\documentclass [12pt] {report}
\usepackage[cp1251]{inputenc}
\usepackage[dvips]{graphicx}
\usepackage{epsfig, latexsym,graphicx}
\usepackage{psfrag}
\usepackage{amsfonts,amsmath,amssymb,amscd,array, euscript}
\newcommand {\eqdef} {\stackrel{\rm def}{=}}

\newcommand {\e} {\mathop{\rm e}\nolimits}

\newcommand {\D}[2] {\displaystyle\frac{\partial{#1}}{\partial{#2}}}

\newcommand {\Dd}[3] {\displaystyle\frac{\partial^2{#1}}{\partial{#2}\partial{#3}}}

\newcommand {\al} {\alpha}

\newcommand {\ga} {\gamma}
\newcommand {\Ga} {\Gamma}

\newcommand {\de} {\delta}
\newcommand {\De} {\Delta}

\newcommand {\fr} {\displaystyle\frac}

\newcommand {\wt} {\widetilde}

\newcommand {\be} {\begin{equation}}
\newcommand {\ee} {\end{equation}}
\newcommand {\ba} {\begin{array}}
\newcommand {\ea} {\end{array}}
\newcommand {\bp} {\begin{picture}}
\newcommand {\ep} {\end{picture}}
\newcommand {\bc} {\begin{center}}
\newcommand {\ec} {\end{center}}

\newcommand {\bt} {\begin{tabular}}
\newcommand {\et} {\end{tabular}}
\newcommand {\lf} {\left}
\newcommand {\rg} {\right}

\newcommand {\nin}{\noindent}

\newcommand {\cA} {{\cal A}}
\newcommand {\cC} {{\cal C}}

\newcommand {\cF} {{\cal F}}

\newcommand {\cI} {{\cal I}}

\newcommand {\br} {{\bf r}}
\newcommand {\bv} {{\bf v}}

\newcommand {\ses} {\medskip}

\def\2#1#2#3{{#1}_{#2}\hspace{0pt}^{#3}}
\def\3#1#2#3#4{{#1}_{#2}\hspace{0pt}^{#3}\hspace{0pt}_{#4}}
\newcounter{sctn}
\def\sec#1.#2\par{\setcounter{sctn}{#1}\setcounter{equation}{0}
                  \noindent{\bf\boldmath#1.#2}\bigskip\par}
\begin {document}

\begin{titlepage}

\begin{center}
{\Large
\bf
 Finslerian metric function of totally anisotropic type.

\ses

 Relativistic aspects
 }\\
\end{center}

\vspace{0.3in}

\begin{center}

\vspace{.15in}
{\large G. S. Asanov\\}
\vspace{.25in}
{\it Division of Theoretical Physics, Moscow State University\\
119992 Moscow, Russia\\
(e-mail: asanov@newmail.ru)}
\vspace{.05in}

\end{center}

\begin{abstract}

The work focuses upon the relativistic and geometric properties
of the space--time
endowed tentatively with the metric function of the Berwald--Moor type.
The zero curvature of indicatrix is a remarkable property of the approach.
We demonstrate how the associated geodesic equations can
 be solved in a transparent way,
thereby obtaining possibility to introduce unambiguously the distance,
angle, and scalar product. We find convenient indicatrix
representation for the associated
tetrads and, by attributing to them naturally the general
meaning of the bases proper of inertial reference frames,
elucidate respective fundamental kinematic relations, including
the extensions of Lorentz transformations and velocity subtraction and composition
laws. The invariance group for the metric tensor is found.

\end{abstract}

\end{titlepage}

\vskip 1cm

\setcounter{sctn}{1} \setcounter{equation}{0}

{\bf 1. Introduction and Motivation
}

\bigskip

The pseudoeuclidean metric function suits the cases when the space-time is uniform in all directions.
Alternatively, we may imagine  a situation when there exist four geometrically distinguished directions
and propose the fundamental metric function
\be
F(y)=\sqrt[4]{|y^1y^2y^3y^4|}
\ee
to measure the length of vectors $y=\{y^A\}$.
By historical reasons (see [1]), the metric function is frequently called after Berwald and Moor
(in any dimension $N\ge3$).

\ses

DEFINITION. Given a centered vector space $V_4$  with some point $``O"$ being the origin
and with the members
$y\in V_4$ issued from the point $``O"$. Let four directions
$\{e_A\}, A=1,2,3,4$
be presupposed in $V_4$.
We may decompose  vectors $y$ with respect to such a basis,
obtaining the component representation $y=\{y^A\}$.
 Under these
conditions, we define the $\cA_4$--{\it space}:
\be
\cA_4~:=\{V_4,e_A,F(y)\}.
 \ee

\ses

According to the known methods of Finsler geometry [1,2],
we construct on the basis of the function $F$
the covariant vector $\hat y=\{y_A\}$ and the Finslerian metric tensor $\{g_{AB}\}$:
\be
 y_A\eqdef\fr12\D{F^2}{y^A}, \qquad
 g_{AB}\eqdef\D{y_A}{y^B}=\fr12\Dd{F^2}{y^A}{y^B}.
\ee

We call the three-dimensional hyperplanes defined by the zeros
$\{y^1=0, ~ y^2=0, ~ y^3=0, ~ y^4=0\}$
of the function $F$ the {\it singular hyperplanes}.
They brake down the space $\cA_4$ into  a collection of 16 sectors, including the
{\it up--sector} $\cA_4^{\{+\}}$.

\ses

DEFINITION. The {\it up--sector} $\cA_4^{\{+\}}\in \cA_4$
is defined by the conditions:
\be
\{
y^1,\,
y^2,\,
y^3,\,
y^4\}
\in
\cA_4^{\{+\}}:~
y^1>0,\,
y^2>0,\,
y^3>0,\,
y^4>0.
\ee
\ses

In what follows, we shall deal with that sector
(unless otherwise is stated explicitly). In such a case the moduli in the right--hand part
of the primary definition (1.1)
can be omitted:
\be
F(y)=\sqrt[4]{y^1y^2y^3y^4}=\sqrt[4]{\prod_{A=1}^4y^A}.
\ee
We shall focus upon the $4$--dimensional case $N=4$; however, many relations and conclusions can be straightforwardly
extended to any dimension $N\ge2$, so that we shall retain in formulae a general $N$ when they are
applicable at arbitrary dimension.


Applying the rules (1.3) to (1.5) yields the explicit component values
\be
y_A=\fr{F^2}{4y^A}
\ee
and
\be
g_{AB}=\fr{2y_Ay_B}{F^2}-\fr{F^2}{4y^Ay^B}\de_{AB};
\ee
the contravariant version of the last tensor is
$\{g^{AB}\}$ with
\be
g^{AB}=\fr{2y^Ay^B}{F^2}-\fr{4y^Ay^B}{F^2}\de^{AB},
\ee
so that
$g^{AC}g_{BC}=\de^A_B$;\, $\de$ stands for the Kronecker symbol.
The time--like nature of the metric tensor
\be
{\rm signature}\{g_{AB}\}=(+ - - -)
\ee
and the constant determinant
\be
\det(g_{AB})=(-1)^{(N+1)}N^{-N}
\ee
(at any dimension number $N$)
are  remarkable properties of the space under study (cf. [1]).
Owing to (1.9), the metric tensor may be represented as
\be
g_{AB}=h^0_Ah^0_B-h^1_Ah^1_B-h^2_Ah^2_B-h^3_Ah^3_B
\ee
in terms of the {\it associated tetrads}
 $\{h_p^A\}$.
The reciprocal representation reads
\be
g^{AB}=h_0^Ah_0^B-h_1^Ah_1^B-h_2^Ah_2^B-h_3^Ah_3^B
\ee
subject to the reciprocity
\be
h_A^ph^A_q=\de^p_q
\ee
(the indices $p,q,\dots$ will be specified over the range $0,1,2,3$ unless otherwise is stated explicitly).
 By comparing (1.10) with (1.11) we may conclude that
\be
\det(h^p_{A})=N^{-N/2}.
\ee


\ses

DEFINITION.
The {\it indicatrix} $\cI_4^{\{+\}}\in\cA_4^{\{+\}}$
is  defined as follows:
\be
y\in \cI_4^{\{+\}}: ~ \{y\in \cA_4^{\{+\}},\,F(y)=1\}.
\ee

\ses

Using the
{\it unit vectors}
\be
l^A\eqdef \fr{y^A}{F(y)}
\ee
(so that $F(l)=1$)
and choosing a convenient parameterization $\{u^a\}, a,b=1,2,3$, over the indicatrix
to have the representation
\be
l^A=t^A(u^a),
\ee
we may construct
the
{\it  projection factors}
\be
t^A_a(u)\eqdef\D{t^A}{u^a}
\ee
for the indicatrix to
obtain the {\it induced metric tensor} over the indicatrix:
\be
i_{ab}(u)
\eqdef -t^A_at^B_bg_{AB};
\ee
here, the minus in front of the right--hand side
reflects the indefinite signature (1.9).
As was demonstrated in [1], is convenient to treat the indicatrix
 in terms of the coordinate
\be
z^0\eqdef \ln F.
\ee


Depending on the sign of the coordinate $z^0$,
the space under study is broken into the unification
\be
\cA_4^{\{+\}}=
\cA^{\{+\}}_{4\,\{z^0>0\}}\cup\cA^{\{+\}}_{4\,\{z^0=0\}}
\cup\cA^{\{+\}}_{4\,\{z^0<0\}}
\ee
of three following regions:
\be
\cA^{\{+\}}_{4\,\{z^0>0\}}~:=\{y\in \cA_4^{\{+\}}:\,F(y)>1\},
\ee
\ses
\be
\cA^{\{+\}}_{4\,\{z^0=0\}}~:=\{y\in \cA_4^{\{+\}}:\,F(y)=1\},
\ee
\ses
\be
\cA^{\{+\}}_{4\,\{z^0<0\}}~:=\{y\in \cA_4^{\{+\}}:\,1>F(y)>0\}.
\ee
Notice that the sector (1.23) is the indicatrix:
\be
\cA^{\{+\}}_{4\,\{z^0=0\}}
=
\cI_4^{\{+\}}.
\ee


The known fact is that if we juxtapose (1.20) by an indicatrix coordinate set $\{u^a\}$ to obtain the four coordinates
\be
z^p=\{z^0,z^a=u^a\},
\ee
then the respective transformation
of the Finslerian metric tensor
would lead to the  result
\be
 g_{AB}(y)\D{y^A}{z^p}\D{y^B}{z^q}=\e^{2z^0}g^*_{pq}
\ee
which is remarkable in that
\be
g^*_{00}=1,\qquad g^*_{0a}=0,\qquad g^*_{ab}=-i_{ab},
\ee
where $\{i_{ab}\}$ is just the indicatrix metric tensor (1.19).
Also, in case of the Finslerian metric function (1.5) the tensor $\{i_{ab}\}$ proves to be exactly euclidean,
so that
the {\it conformal representation}
\be
g^*_{pq}=\e^{2z^0}r_{pq}
\ee
holds with $\{r_{pq}\}$ being the pseudoeuclidean metric tensor.


Therefore, it is attractive to  introduce the {\it associated conformally--pseudoeuclidean space}
$\cC_4$:
\be
\cC_4~:=\{V_4, z^p\in V_4, g^*_{pq}\}
\ee
to have the  {\it isometry}
\be
\cA_4^{\{+\}} \Longleftrightarrow \cC_4
\ee
with the decomposition
\be
\cC_4=\cC_4^{\{+\}}\cup\cC^{\{0\}}\cup \cC_4^{\{-\}},
\ee
where
\be
\cC^{\{+\}}_4~:=\{z^p\in \cC_4^{\{+\}}:\,z^0>0\},
\ee
\ses
\be
\cC^{\{0\}}_4~:=\{z^p\in \cC_4^{\{0\}}:\,z^0=0\},
\ee
\ses
\be
\cC^{\{-\}}_4~:=\{z^p\in \cC_4^{\{-\}}:\,z^0<0\},
\ee
so that
\be
\cA^{\{+\}}_{4\,\{z^0>0\}} \Longleftrightarrow \cC_4^{\{+\}},\qquad
\cA^{\{+\}}_{4\,\{z^0=0\}} \Longleftrightarrow \cC_4^{\{0\}},\qquad
\cA^{\{+\}}_{4\,\{z^0<0\}} \Longleftrightarrow \cC_4^{\{-\}}.
\ee


Now the question is what is the particular and convenient choice for the set
$\{u^a\}$ under which
the tensor
 $\{i_{ab}\}$
 is exactly the diagonal unity, that is, when we get
\be
i_{ab}=\de_{ab}.
\ee
Obviously, in the last case the fundamental length interval $ds$ can be given merely by
\be
(ds)^2=e^{2z^0}\Bigl((dz^0)^2-({\bf dz})^2\Bigr).
\ee
To anticipate  true a due and possible answer to the question, it is useful to note that the choice
\be
l^A=\exp(C^A_au^a)
\ee
with any constant $C^A_a$ subjected to the condition
\be
\sum_{A=1}^4C^A_a=0
\ee
would parameterize the indicatrix because of the  product structure of the Finslerian metric function
(1.5) under study.
Also, if we subject the constants to the condition
\be
\sum_{A=1}^4C^A_aC^B_b=4\de_{ab},
\ee
then, because of the particular structure of the right-hand part in the metric tensor (1.7),
we just obtain $\de_{ab}$ in the right-hand part of (1.19).
When verifying this assertion, it is convenient to note that the projection coefficients (1.18)
constructed on the basis of (1.39) bear the structure
\be
t^A_a=C^A_a\cdot l^A
\ee
at any value of the index $A$.
Owing to the exponential nature of the right-hand part in the representation (1.39), it is convenient to call
the set $\{u^a\}$ the {\it indicatrix variables}.


It is convenient to supplement the constants by the members
\be
C^A_0=1,
\ee
so that
\be
\sum_{A=1}^4C^A_pC^B_q=4e_{pq},
\ee
where $\{e_{pq}\}={\rm diagonal}(1,-1,-1,-1)$ is the pseudoeuclidean metric tensor.
This entails
\be
\sum_{A=1}^4C_A^a=0.
\ee
The inverse constants $C_A^p$ obeying the relations
\be
C^p_AC^A_q=\de^p_q
\ee
must show the properties
\be
C^0_A=\fr14
\ee
and
\be
\sum_{A=1}^4C^A_a=0.
\ee
Under these conditions, the representation (1.39) can be inverted to yield
\be
u^a=C^a_A\ln l^A
\ee
and
\be
z^p=C^p_A\ln l^A,
\ee
which in turn yields for the tetrads
\be
 h^p_A=Fz^p_A=C_A^p\cdot\fr1{l^A},
\ee
where
\be
z^p_A=\D{z^p}{y^A}.
\ee
\ses

From (1.51) it follows that
\be
g_{AB}=F^2c_{AB}
\ee
with the tensor
\be
c_{AB}=z^p_Az^q_Be_{pq},
\ee
which demonstrates that the Finslerian metric tensor associated with the metric function (1.1)
is conformal to the
pseudoeuclidean metric tensor. The conformal multiplier is the square $F^2$ of the metric function $F$.


We shall frequently substitute variables $\{a^A\}$ with $\{y^A\}$:
\be
y^A=a^A,
\ee
thereafter the metric function (1.1) takes on the form
\be
F=\sqrt[4]{a^1a^2a^3a^4}.
\ee

In Section 2 we deal with the  geodesic equations of the space under study.
 It proves possible to find the adequate
explicit solutions thereto in both the initial--value and fixed--edge forms. This opens up the
straightforward way to obtain the angle
\be
\eta(a,b)=
\fr12\fr{F(b)}{F(a)}
\sqrt{
\lf(
\ln
\fr
{
a^1
}{
b^1
}
\rg)^2
+\lf(
\ln
\fr
{
a^2
}{
b^2
}
\rg)^2
+
\lf(
\ln
\fr
{
a^3
}{
b^3
}
\rg)^2
+
\lf(
\ln
\fr
{
a^4
}{
b^4
}
\rg)^2}
\ee
between two vectors by postulating  the cosine theorem.
The angle is actually defined by the unit vectors $l^A(a)$ and $l^a(b)$:
\be
\eta(a,b)=
\fr12
\sqrt{
\lf(
\ln
\fr
{
l^1(a)
}{
l^1(b)
}
\rg)^2
+\lf(
\ln
\fr
{
l^2(a)
}{
l^2(b)
}
\rg)^2
+
\lf(
\ln
\fr
{
l^3(a)
}{
l^3(b)
}
\rg)^2
+
\lf(
\ln
\fr
{
l^4(a)
}{
l^4(b)
}
\rg)^2}
\ee
The associated distance and scalar product are also
found.
The angle is additive
when the vectors point to a fixed geodesic  curve.
In fact, this angle measures the euclidean length in the indicatrix.


In Section 3 we derive step--by--step the kinematic implications of the tetrad choice (1.51).
The kinematic coefficients found are of the unit determinant (see (3.10)). Their structure
(3.21)--(3.25) entail the following
 $\cA_4^{\{+\}}$--{\it kinematic transformations}:
\be
Y'^0=\fr
{Y^0+s^1Y^1+s^2Y^2+s^3Y^3}
{\sqrt[4]{(1+s^1+s^2+s^3)(1-s^1+s^2-s^3)(1+s^1-s^2-s^3)(1-s^1-s^2+s^3)}},
\ee
\ses
\be
Y'^1=\fr
{s^1Y^0+Y^1+s^3Y^2+s^2Y^3}
{\sqrt[4]{(1+s^1+s^2+s^3)(1-s^1+s^2-s^3)(1+s^1-s^2-s^3)(1-s^1-s^2+s^3)}},
\ee
\ses
\be
Y'^2=\fr
{s^2Y^0+s^3Y^1+Y^2+s^1Y^3}
{\sqrt[4]{(1+s^1+s^2+s^3)(1-s^1+s^2-s^3)(1+s^1-s^2-s^3)(1-s^1-s^2+s^3)}},
\ee
\ses
\be
Y'^3=\fr
{s^3Y^0+s^2Y^1+s^1Y^2+Y^3}
{\sqrt[4]{(1+s^1+s^2+s^3)(1-s^1+s^2-s^3)(1+s^1-s^2-s^3)(1-s^1-s^2+s^3)}}
\ee
to extend the ordinary Lorentz transformations,
where $s^1,s^2,s^3$ are components of  three--dimensional motion velocity.
For the velocity,  we obtain the extended composition law as well the subtraction law
in explicit forms.
The  {\it kinematic $\cA^{\{+\}}_4$--length}
\be
\cF(Y)=
\sqrt[4]{(Y^0\!+\!Y^1\!+\!Y^2\!+\!Y^3)
(Y^0\!-\!Y^1\!+\!Y^2\!-\!Y^3)
(Y^0\!+\!Y^1\!-\!Y^2\!-\!Y^3)
(Y^0\!-\!Y^1\!-\!Y^2\!+\!Y^3)}
\ee
is appeared (see (3.30)) which fulfills
the {\it kinematic $\cA^{\{+\}}_4$--invariance}
\be
\cF(Y')=\cF(Y)
\ee
(see (3.31)).
The transformations (1.59)--(1.62)  obviously extend
 the ordinary special--relativistic pseudoeuclidean Lorentz transformations
\be
Y'^0_{\{\rm special~ Lorentzian\}}=\fr
{Y^0+s^1Y^1}
{\sqrt{(1+s^1)(1-s^1)}}, \qquad
Y'^1_{\{\rm special~ Lorentzian\}}=\fr
{s^1Y^0+Y^1}
{\sqrt{(1+s^1)(1-s^1)}},
\ee
accordingly  the result (1.63) extends the function
\be
{\cF(Y)}_{\{\rm special~ Lorentzian\}}=\sqrt{(Y^0+Y^1)(Y^0-Y^1)}.
\ee

In Section 4, we expose  the  transformations that leave invariant the Finslerian metric function as well as
the Finslerian metric tensor.
In the space under study,
the transformations are found to be, in general, nonlinear.
They realize euclidean rotations and translations in the indicatrix. That is to say, the group of such transformations
is a nonlinear image of the euclidean invariance group. The translations in the euclidean indicatrix give rise
to scale (product) transformations in the initial space, so that they form a  linear (and abelian) subroup.
Detailed calculations are presented in Appendices A, B, and C.

The paper ends with a short Discussion of the key aspects of our approach.


\ses\ses

\setcounter{sctn}{2}
 \setcounter{equation}{0}

\nin{\bf 2.
Geodesics, distance, and
 angle in $\cA_N$--spaces}

\ses

Given  a conformally--pseudoeuclidean
space
$\cC_4$  (see (1.30)) with the metric tensor
$\{g^*_{pq}\}$ prescribed by the conformal representation (1.29). Calculating the partial derivatives
$g^*_{pq,r}=\partial g^*_{pq}/\partial z^r$,
we get
$
g^*_{pq,a}=0
$
and
$
g^*_{pq,0}=2g_{pq},
$
so that for the components
$
{\Ga}_{pqr}=\fr12(g^*_{pq,r}+g^*_{qr,p}-g^*_{pq,r})
$
we shall have the values
\be
{\Ga}_{000}=g^*_{00},\quad {\Ga}_{a00}={\Ga}_{0a0}=0,\quad {\Ga}_{a0b}=-g^*_{ab},
\quad {\Ga}_{ab0}=g^*_{ab}, \quad {\Ga}_{abc}=0,\quad {\Ga}_{pq0}=g^*_{pq}.
\ee
The associated Christoffel symbols
$\3{\Ga}prq=g^{*\,rs}{\Ga}_{psq}$
are given by the components
\be
\3{\Ga}000=1, \quad \3{\Ga}a00= \3{\Ga}0a0=0,\quad \3{\Ga}ab0=\de_a^b, \quad \3{\Ga}a0b=r_{ab}, \quad
\3{\Ga}abc=0, \quad\3{\Ga}pq0=\de^q_p.
\ee

Let us consider a curve $C(s)$
parameterized by the length parameter $s$
(cf. (1.38)) and introduce the respective four-dimensional velocity
\be
U^p=\fr{dz^p}{ds},
\ee
 so that the velocity is unit:
\be
g^*_{pq}(z)
U^p
U^q
=1
\ee
(the timelike case).
The differential equation for the  $C(s)$ to be a geodesic curve
\be
\fr{dU^p}{ds}+\3{\Ga}sprU^sU^r=0
\ee
proves to consist of two parts:
\be
\fr {dU^0}{ds}=-
[(U^0)^2+{\bf U}^2]
=-2(U^0)^2+\e^{-2z^0}
\ee
and
\be
\fr {dU^a}{ds}=-
2U^aU^0.
\ee
The equation
\be
\fr{d^2z^0}{ds^2}+
2(\fr{dz^0}{ds})^2=\e^{-2z^0}
\ee
can readily be integrated, yielding
\be
z^0=\ln(f(s))
\ee
with
\be
f(s)=\sqrt{a^2+2bs+s^2},
\ee
where $a$ and $b$ are integration constants.


Since $z^0=\ln F$ (see (1.20)), from (2.10) it follows that
the Finslerian metric function varies along the geodesics according to the law
\be
F(s)=\sqrt{a^2+2bs+s^2}.
\ee
Furthermore,
using
\be
\fr{dz^0}{ds}=\fr{b+s}{(F(s))^2}=U^0
\ee
in (2.7) enables us to readily find
\be
U^a=\fr{
\sqrt{b^2-a^2}
\,
n^a}{(F(s))^2},
\ee
where $n^a$ is a set of constants. To fulfill (2.4), the set must be subjected to the unity length condition:
\be
\de_{ab}n^an^b=1.
\ee
Using $U^a=dz^a/ds$ (see (2.3)) in (2.13) gives us a  differential
equation to find the functions $z^a(s)$.
The  equation can readily be integrated  to yield
\be
z^a(s)=m^a+n^a\fr12\ln
\fr{s+b-
\sqrt{b^2-a^2}}
{s+b+
\sqrt{b^2-a^2}},
\ee
where $m^a$ are new integration constants;
we assume
\be
b^2-a^2\ge 0,\quad a>0.
\ee
Eqs. (2.11)--(2.14) upon the condition (2.16) fulfill  (2.4).

In this way we obtain explicitly the following formulae:
\be
r_1^0=\ln a,\quad r_2^0=\ln(F(\De s)),\quad \sqrt{b^2-a^2}=a^2|\bv_1|, \quad b=a\sqrt{1+a^2|\bv_1|^2},
\ee
\ses
\be
r^0=\fr12\ln(a^2+2bs+s^2),
\ee
\ses
\be
\br(s)=\br_1+\fr12\bv_1\fr{a^2}{\sqrt{b^2-a^2}}\ln\Bigl(X(s)\Bigr),
\ee
where $r^0=z^0$,
$\br=\{z^a_1\}$,
$\bv=\{v^a_1\}$,
and
\be
X(s)=
\fr{s+b-
\sqrt{b^2-a^2}}
{s+b+
\sqrt{b^2-a^2}}
\,
\fr{b+
\sqrt{b^2-a^2}}
{b-
\sqrt{b^2-a^2}}.
\ee
The last function  can also be  represented in the forms
\be
X(s)=
\fr{
\Bigl[
a^2+(b+
\sqrt{b^2-a^2})
 s
\Bigl]^2
}
{a^2F^2(
 s)
}
=
\fr{
\Bigl[
a^2+b
 s
+
\sqrt{b^2-a^2}
\,
 s
\Bigl]^2
}
{a^2F^2(
 s)
}.
\ee

Thus we have arrived at

\ses

PROPOSITION.
The {\it initial--value solution} to the geodesic equations (2.5) under study
can explicitly be given by Eqs. (2.17)--(2.20).

\ses


Also, it is possible to explicate the  representation
\be
{\bf r}(s)={\bf r}_1+\fr{{\bf r}_2-{\bf r}_1}{|{\bf r}_2-{\bf r}_1|
}\ln\sqrt{X(s)},
\ee
with
\be
|{\bf r}_2-{\bf r}_1|=\ln\sqrt{X(\De s)},
\ee
\ses
\be
b=\fr{|{\bf r}_1||{\bf r}_2|\cosh|{\bf r}_2-{\bf r}_1|-|{\bf r}_1|^2}
{\De s},
\ee
\ses
\be
\sqrt{b^2-a^2}=\fr{|{\bf r}_1||{\bf r}_2|\sinh|{\bf r}_2-{\bf r}_1|}
{\De s},
\ee
and
\be
(\De s)^2=|{\bf r}_1|^2+|{\bf r}_2|^2-2|{\bf r}_1||{\bf r}_2|\cosh|{\bf r}_2-{\bf r}_1|.
\ee

Thus, we have obtained

\ses

PROPOSITION. The {\it fixed--edge  solution}
to the geodesic equations (2.5) under study
can explicitly be given by Eqs. (2.22)--(2.26).

\ses

\begin{figure}[!ht]
 \centering
\footnotesize
\psfrag{P1}{$P_1$}
\psfrag{P2}{$P_2$}
\psfrag{PS}{$P(s)$}
\psfrag{C}{$C$}
\psfrag{O}{$O$}
\psfrag{t1}{\hspace{-0.5cm}$a^A(0)$}
\psfrag{t2}{$a^A$}
\psfrag{v1}{$\mathbf{v}_1$}
\psfrag{R1}{$\mathbf{r}_1$}
\psfrag{R2}{$\mathbf{r}_2$}
\psfrag{R3}{$\mathbf{r}(s)$}
\psfrag{U}{$\mathbf{v}_1$}
\psfrag{a}{${\bf t}_1$}
\psfrag{b}{${\bf t}_2$}
\psfrag{e}{$\eta$}
\includegraphics[width=6.cm]{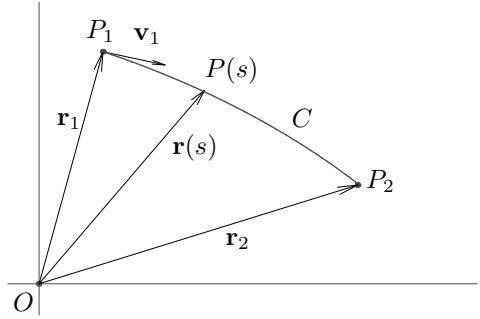}
\caption{\small A geodesic curve $C$; the length of the curve
 from  p. $P_1$ to p. $P_2$ is equal to $\De s$,
and   from  p. $P_1$ to p. $P(s)$  is equal to $s$ }
\end{figure}


The formula (2.26) can also be written as
\be
(\De s)^2=|{\bf r}_1|^2+|{\bf r}_2|^2-2|{\bf r}_1||{\bf r}_2|\cosh\Bigl(\eta({\bf r}_1,{\bf r}_2)\Bigr)
\ee
with the following $\cC_4${\it --angle}:
\be
\eta({\bf r}_1,{\bf r}_2)=
|{\bf r}_2-{\bf r}_1|.
\ee

\ses

PROPOSITION.
The $\cC_4${\it --cosine theorem} reads as (2.26) or (2.28).

\ses

In view of (2.21) and (2.24)--(2.25), we can write
\be
X(\De s)=\e^{2\eta}.
\ee

\ses\ses

\begin{figure}[!ht]
 \centering
\footnotesize
\psfrag{P1}{$P_1$}
\psfrag{P2}{$P_2$}
\psfrag{PS}{$P(s)$}
\psfrag{C}{$C$}
\psfrag{O}{$O$}
\psfrag{t1}{\hspace{-0.5cm}$a^A(0)$}
\psfrag{t2}{$a^A$}
\psfrag{v1}{$\mathbf{v}_1$}
\psfrag{R1}{$a^A_{\{1\}}$}
\psfrag{R2}{$a^A_{\{2\}}$}
\psfrag{R3}{$a^A(s)$}
\psfrag{U}{$\mathbf{v}_1$}
\psfrag{1}{}\psfrag{2}{$$}\psfrag{3}{}\psfrag{4}{}\psfrag{6}{}\psfrag{-1}{}\psfrag{-}{}
\psfrag{a}{${\bf  r}_1$}
\psfrag{b}{${\bf r}_2$}
\psfrag{e}{$\eta$}
\includegraphics[width=7.cm]{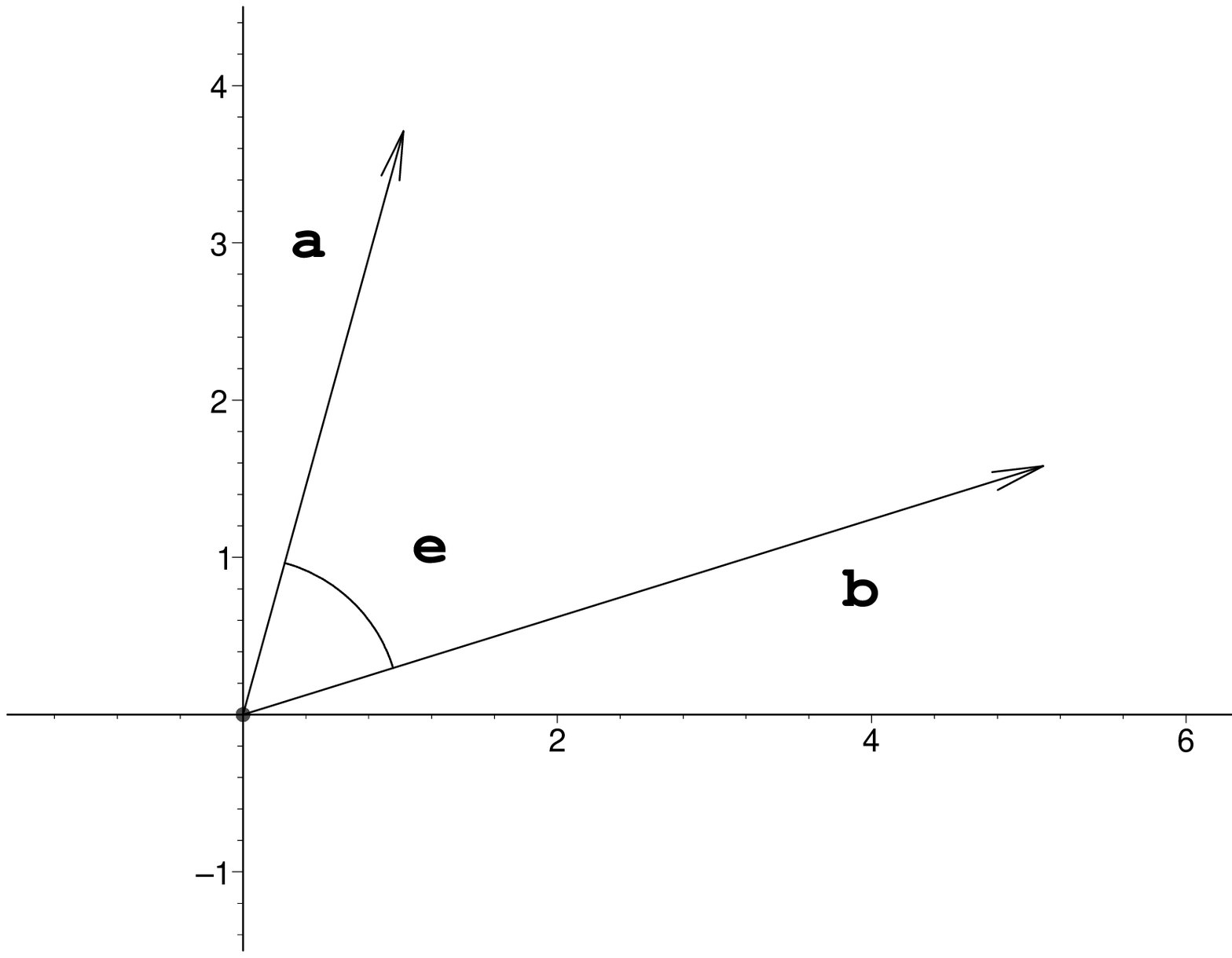}
\caption{\small The angle $\eta(\br_1,\br_2)$}
\end{figure}


By comparing (2.15) and (2.20) with the unit vector representation of the exponential type (1.39),
we can readily conclude that
 the components of the unit vector $l^A$ vary
  along geodesics in accordance with the law
\be
l^A(s)=
l^A(0)
\Bigl(X(s)
\Bigr)^
{\ln_{X(\De s)}(
l^A(\De s)/
l^A(0)
)
},
\ee
where
\be
\prod_{A=1}^N{l^A(0)}=1,
\qquad
\prod_{A=1}^N{l^A(\De s)}=1,
\qquad
\prod_{A=1}^N{l^A(s)}=1.
\ee
This law  is applicable at any dimension $N\ge2$.

For the vector
\be
a^A(s)=F(s)l^A(s)
\ee
we obtain from (2.30) the similar behaviour
\be
a^A(s)=
\fr{F(s)a^A(0)}{F(0)}
\Bigl(X(s)
\Bigr)^
{\ln_{X(\De s)}(
a^A(\De s)F(0)/
a^A(0)F(\De s)
)
},
\ee
where $F(0)=a$ (in view of (2.11)).

Thus we have arrived at

\ses

PROPOSITION.
Given two vectors $\{a^A_{\{1\}}\}$ and  $\{a^A_{\{2\}}\}$.
 Let $C$ be a  curve going from the end of the first vector to the end of the second vector.
Put $a^A(0)=\{a^A_{\{1\}}\}$ and  $a^A(\De s)=\{a^A_{\{2\}}\}$.
Attribute the length values $s=0$ and $s= \De s$ to the vectors,
where $\De s$ is the length of the curve $C$.
 If $C$ is a geodesics, then the vector stretching to the geodesics point with a length value $s$
is explicitly given by (2.33).

\ses

\ses

\begin{figure}[!ht]
 \centering
\footnotesize
\psfrag{P1}{$P_1$}
\psfrag{P2}{$P_2$}
\psfrag{PS}{$P(s)$}
\psfrag{C}{$C$}
\psfrag{O}{$O$}
\psfrag{t1}{\hspace{-0.5cm}$a^A(0)$}
\psfrag{t2}{$a^A$}
\psfrag{v1}{$\mathbf{v}_1$}
\psfrag{R1}{$a^A_{\{1\}}$}
\psfrag{R2}{$a^A_{\{2\}}$}
\psfrag{R3}{$a^A(s)$}
\psfrag{U}{$\mathbf{v}_1$}
\psfrag{a}{${\bf t}_1$}
\psfrag{b}{${\bf t}_2$}
\psfrag{e}{$\eta$}
\includegraphics[width=6.cm]{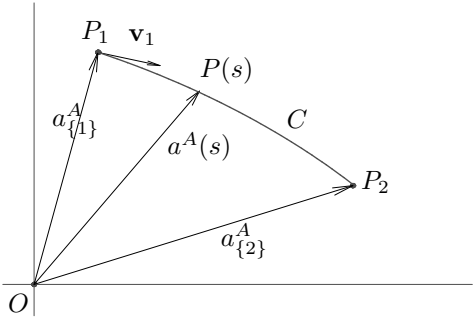}
\caption{\small The solution to the geodesic equation in the $\cA_4^{\{+\}}$}
\end{figure}


The result (2.11) entails the relation
\be
a^2F^2(
\De s)
=
(a^2+b
\De s
)^2
-
(\sqrt{b^2-a^2}
\,
\De s
)^2,
\ee
which can be used to introduce the angle $\eta$ according to
\be
a^2+b
\De s
=
aF(
\De s)
\cosh\eta
\ee
and
\be
\sqrt{b^2-a^2}
\,
\De s
=
aF(
\De s)
\sinh\eta,
\ee
or
\be
\fr{a^2+b
\De s
}
{\sqrt{b^2-a^2}
\,
\De s
}
=
\tanh\eta.
\ee
Applying (2.35) and (2.36) to (2.21),
it follows that
\be
X(\De s)=(\cosh\eta+\sinh\eta)^2=\e^{2\eta},
\ee
so that the equality (2.29) has been reproduced.
Therefore,
we may write
the laws (2.30) and (2.33) in the forms
\be
l^A(s)=
l^A(0)
\Bigl(
X(s)
\Bigr)^
{\fr1{2\eta}
\ln(
l^A(\De s)/
l^A(0)
)
}
\ee
and
\be
a^A(s)=
\fr{F(s)a^A(0)}{F(0)}
\Bigl(
X(s)
\Bigr)^
{\fr1{2\eta}
\ln(
a^A(\De s)F(0)/
a^A(0)F(\De s)
)
}.
\ee


Since
\be
\fr{d\Bigl(
\fr{s+b-
\sqrt{b^2-a^2}}
{s+b+
\sqrt{b^2-a^2}}\Bigr)}
{ds}
=
\fr{2\sqrt{b^2-a^2}}{F^2}
\fr{s+b-
\sqrt{b^2-a^2}}
{s+b+
\sqrt{b^2-a^2}},
\ee
from (2.16) and (2.40) we can conclude that
\be
F(s)\fr{da^A}{ds}=\fr{dF(s)}{ds}a^A(s)+
2\sqrt{b^2-a^2}
\,
a^A(s)
\fr1{2\eta}
\ln(
l^A(\De s)/
l^A(0)
).
\ee
Using here
\be
g_{AB}=\fr{2a_Aa_B}{F^2}-\fr{F^2}{Na^Aa^B}\de_{AB},
\qquad
a_A=\fr{F^2}{Na^A}
\ee
(see (1.6) and (1.7)),
and noting
that
\be
\prod_{A=1}^N
\ln(
l^A(\De s)/
l^A(0)
)=0,
\ee
we find
the equality
$$
g_{AB}(a^C)
\fr{da^A}{ds}
\fr{da^B}{ds}
=(
\fr{dF}{ds})^2
-
(b^2-a^2)
\fr1{F^2}
\fr1{N\eta^2}
\sum_{A=1}^N
\lf(
\ln(
l^A(\De s)/
l^A(0)
)
\rg)^2
$$
\ses
\be
=1+\fr{b^2-a^2}{F^2}
-
(b^2-a^2)
\fr1{F^2}
\fr1{N\eta^2}
\sum_{A=1}^N
\lf(
\ln(
l^A(\De s)/
l^A(0)
)
\rg)^2.
\ee
The left-hand side here
must be $1$. Therefore, the angle $\eta$ can be given by
\be
\eta=
\sqrt{\fr1N
\sum_{A=1}^N
\lf(
\ln
\fr
{
l^A(\De s)
}{
l^A(0)
}
\rg)^2
},
\ee
or equivalently,
\be
\eta=
\sqrt{\fr1N
\sum_{A=1}^N
\lf(
\ln
\fr
{
a^A(\De s)F(0)
}{
a^A(0)F(\De s)
}
\rg)^2
}.
\ee

If we merely consider two vectors $\{a^A\}$ and $\{b^A\}$, then (2.47) assigns for them the respective angle
\be
\eta(a,b)=
\sqrt{\fr1N
\sum_{A=1}^N
\lf(
\ln
\fr
{
a^AF(b)
}{
b^AF(a)
}
\rg)^2
}.
\ee


Thus, the following assertions are valid.

\ses

PROPOSITION. The angle between two vectors
$\{a^A\}$ and $\{b^A\}$
is given by (2.48).
The angle is symmetric
\be
\eta(a,b)=\eta(b,a).
\ee
Also,
the angle is additive
\be
\eta(a,b)+\eta(b,c)
=
\eta(a,c),
\ee
when the vectors $\{a^A,b^A,c^A\}$ point to a fixed geodesic  curve.
\ses

\ses

\begin{figure}[!ht]
 \centering
\footnotesize
\psfrag{P1}{$P_1$}
\psfrag{P2}{$P_2$}
\psfrag{PS}{$P(s)$}
\psfrag{C}{$C$}
\psfrag{O}{$O$}
\psfrag{t1}{\hspace{-0.5cm}$a^A(0)$}
\psfrag{t2}{$a^A$}
\psfrag{v1}{$\mathbf{v}_1$}
\psfrag{R1}{$a^A_{\{1\}}$}
\psfrag{R2}{$a^A_{\{2\}}$}
\psfrag{R3}{$a^A(s)$}
\psfrag{U}{$\mathbf{v}_1$}
\psfrag{1}{}\psfrag{2}{$$}\psfrag{3}{}\psfrag{4}{}\psfrag{6}{}\psfrag{-1}{}
\psfrag{-}{}
\psfrag{a}{$a^A_{\{1\}}$}
\psfrag{b}{$a^A_{\{2\}}$}
\psfrag{e}{$\eta$}
\includegraphics[width=7.cm]{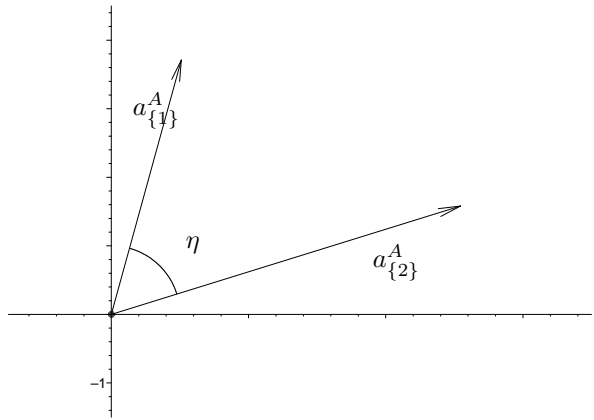}
\caption{\small The angle $\eta(a_{\{1\}},a_{\{2\}})$}
\end{figure}


Rewriting (2.34) as
\be
F^2(\Delta s)=
(\Delta s)^2-a^2+
2(a^2+b
\Delta s
)
\ee
and use (2.35), we get

\ses

\nin
the {\it Finslerian $\cA_N^{\{+\}}$--Cosine Theorem:}
\ses\\
\be
(\Delta s)^2
=
(F(a))^2+(F(b))^2
-2
F(a)F(b)
\cosh\Bigl(\eta(a,b)\Bigr).
\ee

Therefore,
the {\it Finslerian $\cA_N^{\{+\}}$--Distance}
$|b\ominus a|$
 between  end points of two given vectors is
\be
|b\ominus a|^2=
(F(a))^2+(F(b))^2
-2
F(a)F(b)
\cosh\Bigl(\eta(a,b)\Bigr).
\ee
The {\it Finslerian $\cA_N^{\{+\}}$--Scalar Product}
\be
(ab)=
F(a)F(b)
\cosh\Bigl(\eta(a,b)\Bigr)
\ee
is obtained.


\ses

NOTE. In the dimension
\be
N=4
\ee
we may  use in the above expression
(2.46)  the indicatrix representation (1.39)
and apply (1.40).
On so doing, we obtain
\be
\eta
 =\sqrt{
 (\De u^1)^2+(\De u^2)^2
+
(\De u^3)^2}.
\ee
Since at the same time
the variables $\{u^1,u^2,u^3\}$ are some euclidean coordinates on the indicatrix (see (1.37)), we may state the following
result.

\ses

PROPOSITION. The Finslerian angle  $\eta$ is tantamount to the indicatrix euclidean distance.

\ses

It may also be said that, to entire analogy to the euclidean geometry proper,
the {\it Finslerian angle $\eta$ found measures the geodesic lengths on the incicatrix.}
However, in the euclidean geometry the arcs are pieces of circles (the euclidean indicatrix is a unit sphere),
 while in our
present case they are pieces of straightlines (since the indicatrix is a euclidean plane).
It is useful to compare (2.56) with the representation (2.28) of the angle $\eta$.


\ses

NOTE.
The two--dimensional case
\be
N=2
\ee
is also comprised by the above formulae.
Namely, the  $\{N=2\}$--dimensional precursor to the angle (2.48) reads
$$
\eta_{\{N=2\}}(a,b)=
\sqrt{\fr12
\sum_{A=1}^2
\lf(
\ln
\fr
{
a^AF(b)
}{
b^AF(a)
}
\rg)^2
}
=
\sqrt{\fr12
\Biggl(
\lf(
\ln
\fr
{
a^1F(b)
}{
b^1F(a)
}
\rg)^2
+
\lf(
\ln
\fr
{
a^2F(b)
}{
b^2F(a)
}
\rg)^2
}\Biggr)
$$
(with $F(a)=\sqrt{a^1a^2}$ and $F(b)=\sqrt{b^1b^2})$,
or
\be
\eta_{\{N=2\}}(a,b)=
\ln
\fr
{
a^1F(b)
}{
b^1F(a)
}.
\ee
Therefore,
\be
\cosh(\eta_{\{N=2\}}(a,b))
=\fr12\Bigl(
\fr
{
a^1F(b)
}{
b^1F(a)
}
+\fr
{
b^1F(a)
}{
a^1F(b)
}
\Bigr)
=\fr{a^1b^2+b^1a^2}{F(a)F(b)}.
\ee
On taking
\be
m^1=\fr{a^1+a^2}2,\quad m^2=\fr{a^1-a^2}2,
\qquad
n^1=\fr{b^1+b^2}2,\quad n^2=\fr{b^1-b^2}2,
\ee
our explication(2.59) just reduces to the ordinary relativistic rule
\be
\cosh(\eta_{\{N=2\}}(a,b))
=\fr{m^1n^1-m^2n^2}{\sqrt{(m^1)^2-(m^2)^2}\,\sqrt{(n^1)^2-(n^2)^2}}.
\ee

\ses


\setcounter{sctn}{3} \setcounter{equation}{0}

{\bf 3. Explicated extension of Lorentzian relations}

\ses\ses

To treat the kinematic topics, we are to consider two (inertial) reference frames $S\{a\}$ and $S\{b\}$
moving in the four--dimensional directions of vectors $a^A$ and $b^A$.
The tetrads $h^p_A(a)$ and $h^p_A(b)$ play naturally the roles of their reference systems proper.
Let a signal move in the direction of a four--dimensional vector $R^A$.
Then with respect to the frames the components of the vector are
\be
R^p_{\{a\}}=h^p_A(a)R^A,
\qquad
R^p_{\{b\}}=h^p_A(b)R^A,
\ee
respectively.
Therefore,
the transformation law
\be
R^p_{\{a\}}=N^p_q(a,b)R^q_{\{b\}}
\ee
 from the reference frame  $S\{b\}$ into the reference frame  $S\{a\}$
is realized by means of the {\it kinematic coefficients}
\be
N^p_q(a,b)=h^p_A(a)h^A_q(b).
\ee
If we apply here  the representation
(1.50)--(1.52)
for the tetrads, we find that the very coefficients
$C_A^p$ disappear in the right--hand part of (3.3), the {\it symmetry}
\be
N^p_q=N^q_p,
\ee
holds,
and
the components of (3.3)
are given explicitly by means of the formulae
\be
N^1_2=N^3_0,\qquad N^1_3=N_0^2,\qquad N^2_3=N^1_0,
\ee
and
\be
N^0_0(a,b)=\fr{F(a)}{4F(b)}
\Bigl(
\fr{b^1}{a^1}+
\fr{b^2}{a^2}+\fr{b^3}{a^3}+\fr{b^4}{a^4}
\Bigr),
\ee
\ses
\be
N^1_0(a,b)=\fr{F(a)}{4F(b)}
\Bigl(
\fr{b^1}{a^1}-
\fr{b^2}{a^2}+\fr{b^3}{a^3}-\fr{b^4}{a^4}
\Bigr),
\ee
\ses\be
N^2_0(a,b)=\fr{F(a)}{4F(b)}
\Bigl(
\fr{b^1}{a^1}+
\fr{b^2}{a^2}-\fr{b^3}{a^3}-\fr{b^4}{a^4}
\Bigr),
\ee
\ses\be
N^3_0(a,b)=\fr{F(a)}{4F(b)}
\Bigl(
\fr{b^1}{a^1}-
\fr{b^2}{a^2}-\fr{b^3}{a^3}+\fr{b^4}{a^4}
\Bigr),
\ee
from which it follows that the determinant is unit
\be
\det(N^p_q)=1,
\ee
the identity
\be
(N^0_0+N^1_0+N^2_0+N^3_0)
(N^0_0-N^1_0+N^2_0-N^3_0)
(N^0_0+N^1_0-N^2_0-N^3_0)
(N^0_0-N^1_0-N^2_0+N^3_0)=1
\ee
takes place,
and


\ses

PROPOSITION.  The {\it group property}
\be
N^p_q(a,c)=
N^p_r(a,b)
N^r_q(b,c)
\ee
holds.

\ses

To verify the identity (3.11) it is worth evaluating from (3.6)--(3.9) the equalities
\be
\fr{F(a)}{F(b)}\fr{b^1}{a^1}
=N^0_0(a,b)+N^1_0(a,b)+N^2_0(a,b)+N^3_0(a,b),
\ee
\ses
\be
\fr{F(a)}{F(b)}\fr{b^2}{a^2}=N^0_0(a,b)-N^1_0(a,b)+N^2_0(a,b)-N^3_0(a,b),
\ee
\ses
\be
\fr{F(a)}{F(b)}\fr{b^3}{a^3}=N^0_0(a,b)+N^1_0(a,b)-N^2_0(a,b)-N^3_0(a,b),
\ee
\ses
\be
\fr{F(a)}{F(b)}\fr{b^4}{a^4}=N^0_0(a,b)-N^1_0(a,b)-N^2_0(a,b)+N^3_0(a,b).
\ee
\ses


The {\it three-dimensional relative velocity}
\be
s^a(a,b)=\fr{N^a_0(a,b)}{N^0_0(a,b)}\equiv \fr{h^a_A(a)b^A}{h_B^0(a)b^B}
\ee
of the reference frame  $S\{b\}$ with respect to the reference frame  $S\{a\}$,
together with the adjoint velocity measure
\be
s_a(a,b)=\fr{N_a^0(a,b)}{N_0^0(a,b)},
\ee
can naturally be introduced.
It proves that
\be
s^a=s_a.
\ee
The kinematic coefficients can be written as functions of the velocity $s^a$:
\be
N^p_q=N^p_q(s^a).
\ee
Indeed, using (3.11) and introducing the {\it relativistic $\cA^{\{+\}}_4$--dilatation factor}
\be
A(s)=
\sqrt[4]{(1+s^1+s^2+s^3)(1-s^1+s^2-s^3)(1+s^1-s^2-s^3)(1-s^1-s^2+s^3)},
\ee
we can establish the equalities
\be
N^0_0=N^1_1=N^2_2=N^3_3=\fr1A,
\ee
\ses
\be
 N^a_0=\fr1As^a,
\ee
\ses
\be
N^0_a=\fr1As_a,
\ee
\ses
\be
N^1_2=N^2_1=\fr1A{s^3},
\quad
N^1_3=N^3_1=\fr1A{s^2},
\quad
N^2_3=N^3_2=\fr1A{s^1}.
\ee

Thus, the following result can be formulated.

\ses

PROPOSITION.  The kinematic coefficients depend on the vectors  $a^A$ and $b^A$ through only the relative
velocity $s^a$.

\ses

This assertion can be meant to claim  the {\it  extended relativity principle}.

For  various purposes of calculations it occurs convenient to rewrite (3.13)--(3.16) as follows:
\be
 \fr{F(a)}{F(b)}\fr{b^1}{a^1}
=\fr1{A(s)}(1+s^1+s^2+s^3),
\ee
 \ses
\be
 \fr{F(a)}{F(b)}\fr{b^2}{a^2}
=\fr1{A(s)}(1-s^1+s^2-s^3),
\ee
 \ses
 \be
 \fr{F(a)}{F(b)}\fr{b^3}{a^3}
=\fr1{A(s)}(1+s^1-s^2-s^3),
\ee
 \ses
 \be
 \fr{F(a)}{F(b)}\fr{b^4}{a^4}
=\fr1{A(s)}(1-s^1-s^2+s^3).
\ee


The above observations suggest the idea to introduce

\ses

DEFINITION. Given a four--dimensional vector
$Y=\{Y^p\}$
in an inertial reference frame.
The {\it kinematic $\cA^{\{+\}}_4$--length}
$\cF(Y)$ of the vector  reads
\be
\cF(Y)=
\sqrt[4]{(Y^0\!+\!Y^1\!+\!Y^2\!+\!Y^3)
(Y^0\!-\!Y^1\!+\!Y^2\!-\!Y^3)
(Y^0\!+\!Y^1\!-\!Y^2\!-\!Y^3)
(Y^0\!-\!Y^1\!-\!Y^2\!+\!Y^3)}.
\ee

The definition entails
the {\it $\cA^{\{+\}}_4$--invariance}
\be
\cF(N^p_q(s^a)Y^q)=\cF(Y^p).
\ee

There exists a simple way to verify the  invariance.
Namely,
applying the coefficients (3.22)--(3.25) to the parentheses appeared under the
root in the right--hand part of (3.30) yields
\be
(Y^0\!+\!Y^1\!+\!Y^2\!+\!Y^3)\to
\fr1{A(s)}(1+s^1+s^2+s^3)(Y^0\!+\!Y^1\!+\!Y^2\!+\!Y^3),
\ee
\be
(Y^0\!-\!Y^1\!+\!Y^2\!-\!Y^3)\to
\fr1{A(s)}(1-s^1+s^2-s^3)(Y^0\!-\!Y^1\!+\!Y^2\!-\!Y^3),
\ee
\be
(Y^0\!+\!Y^1\!-\!Y^2\!-\!Y^3)\to
\fr1{A(s)}(1+s^1-s^2-s^3)(Y^0\!+\!Y^1\!-\!Y^2\!-\!Y^3),
\ee
\be
(Y^0\!-\!Y^1\!-\!Y^2\!+\!Y^3)\to
\fr1{A(s)}(1-s^1-s^2+s^3)(Y^0\!-\!Y^1\!-\!Y^2\!+\!Y^3),
\ee
so that  on taking into account the right--hand part in the expression
(3.21)
of  $ A(s) $
we may just establish (3.31)
(that is to say, the quantities $s^a$ do  disappear in the left--hand part of (3.31)).


Let us find the subtraction  law
\be
s_{\{1\}}=s_{\{3\}}\ominus s_{\{2\}}
\ee
 for the velocities introduced.
The law is an explication from the group law (3.12).
Given  three vectors
$a^A,b^A,c^A$,
we may consider the velocities
$
s^a_{\{1\}}=s^a_{\{1\}}(a,b),~s^a_{\{2\}}=s^a_{\{2\}}(b,c),~s^a_{\{3\}}=s^a_{\{3\}}(a,c)
$
and
find the components
\be
s^1_{\{1\}}=
\fr
{
\fr{b^1}{a^1}-
\fr{b^2}{a^2}+\fr{b^3}{a^3}-\fr{b^4}{a^4}}
{
\fr{b^1}{a^1}+
\fr{b^2}{a^2}+\fr{b^3}{a^3}+\fr{b^4}{a^4}},
\ee
\ses
\be
s^2_{\{1\}}=
\fr
{
\fr{b^1}{a^1}+
\fr{b^2}{a^2}-\fr{b^3}{a^3}-\fr{b^4}{a^4}}
{
\fr{b^1}{a^1}+
\fr{b^2}{a^2}+\fr{b^3}{a^3}+\fr{b^4}{a^4}},
\ee
\ses
\be
s^3_{\{1\}}=
\fr
{
\fr{b^1}{a^1}-
\fr{b^2}{a^2}-\fr{b^3}{a^3}+\fr{b^4}{a^4}}
{
\fr{b^1}{a^1}+
\fr{b^2}{a^2}+\fr{b^3}{a^3}+\fr{b^4}{a^4}},
\ee
which can be transformed to read
\be
s^1_{\{1\}}=
\fr
{
\fr{b^1}{c^1}\fr{c^1}{a^1}-
\fr{b^2}{c^2}\fr{c^2}{a^2}+\fr{b^3}{c^3}\fr{c^3}{a^3}-\fr{b^4}{c^4}\fr{c^4}{a^4}
}
{
\fr{b^1}{c^1}\fr{c^1}{a^1}+
\fr{b^2}{c^2}\fr{c^2}{a^2}+\fr{b^3}{c^3}\fr{c^3}{a^3}+\fr{b^4}{c^4}\fr{c^4}{a^4}
},
\ee
\ses
\be
s^2_{\{1\}}=
\fr
{
\fr{b^1}{c^1}\fr{c^1}{a^1}+
\fr{b^2}{c^2}\fr{c^2}{a^2}-\fr{b^3}{c^3}\fr{c^3}{a^3}-\fr{b^4}{c^4}\fr{c^4}{a^4}
}
{
\fr{b^1}{c^1}\fr{c^1}{a^1}+
\fr{b^2}{c^2}\fr{c^2}{a^2}+\fr{b^3}{c^3}\fr{c^3}{a^3}+\fr{b^4}{c^4}\fr{c^4}{a^4}
},
\ee
\ses
\be
s^3_{\{1\}}=
\fr
{
\fr{b^1}{c^1}\fr{c^1}{a^1}-
\fr{b^2}{c^2}\fr{c^2}{a^2}-\fr{b^3}{c^3}\fr{c^3}{a^3}+\fr{b^4}{c^4}\fr{c^4}{a^4}
}
{
\fr{b^1}{c^1}\fr{c^1}{a^1}+
\fr{b^2}{c^2}\fr{c^2}{a^2}+\fr{b^3}{c^3}\fr{c^3}{a^3}+\fr{b^4}{c^4}\fr{c^4}{a^4}
}.
\ee


We may apply here (3.26)--(3.29) in terms of the quantities
\be
J_{11}=1+s_{\{1\}}^1+s_{\{1\}}^2+s_{\{1\}}^3,
\qquad
J_{21}=1+s_{\{2\}}^1+s_{\{2\}}^2+s_{\{2\}}^3,
\qquad
J_{31}=1+s_{\{3\}}^1+s_{\{3\}}^2+s_{\{3\}}^3,
\ee
\ses
\be
J_{12}=1-s_{\{1\}}^1+s_{\{1\}}^2-s_{\{1\}}^3,
\qquad
J_{22}=1-s_{\{2\}}^1+s_{\{2\}}^2-s_{\{2\}}^3,
\qquad
J_{32}=1-s_{\{3\}}^1+s_{\{3\}}^2-s_{\{3\}}^3,
\ee
\ses
\be
J_{13}=1+s_{\{1\}}^1-s_{\{1\}}^2-s_{\{1\}}^3,
\qquad
J_{23}=1+s_{\{2\}}^1-s_{\{2\}}^2-s_{\{2\}}^3,
\qquad
J_{33}=1+s_{\{3\}}^1-s_{\{3\}}^2-s_{\{3\}}^3,
\ee
\ses
\be
J_{14}=1-s_{\{1\}}^1-s_{\{1\}}^2+s_{\{1\}}^3,
\qquad
J_{24}=1-s_{\{2\}}^1-s_{\{2\}}^2+s_{\{2\}}^3,
\qquad
J_{34}=1-s_{\{3\}}^1-s_{\{3\}}^2+s_{\{3\}}^3,
\ee
obtaining
\be
s^1_{\{3\}}=
\fr
{J_{31}/J_{21}-J_{32}/J_{22}+J_{33}/J_{23}-J_{34}/J_{24}}
{J_{31}/J_{21}+J_{32}/J_{22}+J_{33}/J_{23}+J_{34}/J_{24}},
\ee
\ses
\be
s^2_{\{3\}}=
\fr
{J_{31}/J_{21}+J_{32}/J_{22}-J_{33}/J_{23}-J_{34}/J_{24}}
{J_{31}/J_{21}+J_{32}/J_{22}+J_{33}/J_{23}+J_{34}/J_{24}},
\ee
\ses
\be
s^3_{\{3\}}=
\fr
{J_{31}/J_{21}-J_{32}/J_{22}-J_{33}/J_{23}+J_{34}/J_{24}}
{J_{31}/J_{21}+J_{32}/J_{22}+J_{33}/J_{23}+J_{34}/J_{24}}.
\ee
Simplifying  the right--hand parts eventually gives


\be
Hs^1_{\{1\}}\!=\!
\fr{1\!+\!s^1_{\{3\}}\!+\!s^2_{\{3\}}\!+\!s^3_{\{3\}}}
{1\!+\!s^1_{\{2\}}\!+\!s^2_{\{2\}}+\!s^3_{\{2\}}}
-
\fr{1\!-\!s^1_{\{3\}}\!+\!s^2_{\{3\}}\!-\!s^3_{\{3\}}}
{1\!-\!s^1_{\{2\}}\!+\!s^2_{\{2\}}\!-\!s^3_{\{2\}}}
+
\fr{1\!+\!s^1_{\{3\}}\!-\!s^2_{\{3\}}\!-\!s^3_{\{3\}}}
{1\!+\!s^1_{\{2\}}\!-\!s^2_{\{2\}}\!-\!s^3_{\{2\}}}
-\fr
{1\!-\!s^1_{\{3\}}\!-\!s^2_{\{3\}}\!+\!s^3_{\{3\}}}
{1\!-\!s^1_{\{2\}}\!-\!s^2_{\{2\}}\!+\!s^3_{\{2\}}},
\ee
\ses
\be
Hs^2_{\{1\}}\!=\!
\fr{1\!+\!s^1_{\{3\}}\!+\!s^2_{\{3\}}\!+\!s^3_{\{3\}}}
{1\!+\!s^1_{\{2\}}\!+\!s^2_{\{2\}}+\!s^3_{\{2\}}}
+
\fr{1\!-\!s^1_{\{3\}}\!+\!s^2_{\{3\}}\!-\!s^3_{\{3\}}}
{1\!-\!s^1_{\{2\}}\!+\!s^2_{\{2\}}\!-\!s^3_{\{2\}}}
-
\fr{1\!+\!s^1_{\{3\}}\!-\!s^2_{\{3\}}\!-\!s^3_{\{3\}}}
{1\!+\!s^1_{\{2\}}\!-\!s^2_{\{2\}}\!-\!s^3_{\{2\}}}
-\fr
{1\!-\!s^1_{\{3\}}\!-\!s^2_{\{3\}}\!+\!s^3_{\{3\}}}
{1\!-\!s^1_{\{2\}}\!-\!s^2_{\{2\}}\!+\!s^3_{\{2\}}},
\ee
\ses
\be
Hs^3_{\{1\}}\!=\!
\fr{1\!+\!s^1_{\{3\}}\!+\!s^2_{\{3\}}\!+\!s^3_{\{3\}}}
{1\!+\!s^1_{\{2\}}\!+\!s^2_{\{2\}}+\!s^3_{\{2\}}}
-
\fr{1\!-\!s^1_{\{3\}}\!+\!s^2_{\{3\}}\!-\!s^3_{\{3\}}}
{1\!-\!s^1_{\{2\}}\!+\!s^2_{\{2\}}\!-\!s^3_{\{2\}}}
-
\fr{1\!+\!s^1_{\{3\}}\!-\!s^2_{\{3\}}\!-\!s^3_{\{3\}}}
{1\!+\!s^1_{\{2\}}\!-\!s^2_{\{2\}}\!-\!s^3_{\{2\}}}
+\fr
{1\!-\!s^1_{\{3\}}\!-\!s^2_{\{3\}}\!+\!s^3_{\{3\}}}
{1\!-\!s^1_{\{2\}}\!-\!s^2_{\{2\}}\!+\!s^3_{\{2\}}},
\ee
\ses\ses
with
\be
H\!=\!
\fr{1\!+\!s^1_{\{3\}}\!+\!s^2_{\{3\}}\!+\!s^3_{\{3\}}}
{1\!+\!s^1_{\{2\}}\!+\!s^2_{\{2\}}+\!s^3_{\{2\}}}
+
\fr{1\!-\!s^1_{\{3\}}\!+\!s^2_{\{3\}}\!-\!s^3_{\{3\}}}
{1\!-\!s^1_{\{2\}}\!+\!s^2_{\{2\}}\!-\!s^3_{\{2\}}}
+
\fr{1\!+\!s^1_{\{3\}}\!-\!s^2_{\{3\}}\!-\!s^3_{\{3\}}}
{1\!+\!s^1_{\{2\}}\!-\!s^2_{\{2\}}\!-\!s^3_{\{2\}}}
+\fr
{1\!-\!s^1_{\{3\}}\!-\!s^2_{\{3\}}\!+\!s^3_{\{3\}}}
{1\!-\!s^1_{\{2\}}\!-\!s^2_{\{2\}}\!+\!s^3_{\{2\}}}.
\ee

Thus we have arrived at the following result.

\ses

PROPOSITION.
Under the Finslerian treatment of the $\cA^{\{+\}}_4$--space,
the {\it law of subtraction} (3.36)
 for three-dimensional relativistic  velocities
  is given by the
explicit formulas
(3.50)--(3.53).

\ses


The composition law
\be
s_{\{3\}}=s_{\{1\}}\oplus s_{\{2\}}
\ee
can be evaluated in a similar fashion.
Namely, we have
\be
s^1_{\{3\}}=
\fr
{
\fr{c^1}{a^1}-
\fr{c^2}{a^2}+\fr{c^3}{a^3}-\fr{c^4}{a^4}}
{
\fr{c^1}{a^1}+
\fr{c^2}{a^2}+\fr{c^3}{a^3}+\fr{c^4}{a^4}},
\ee
\ses
\be
s^2_{\{3\}}=
\fr
{
\fr{c^1}{a^1}+
\fr{c^2}{a^2}-\fr{c^3}{a^3}-\fr{c^4}{a^4}}
{
\fr{c^1}{a^1}+
\fr{c^2}{a^2}+\fr{c^3}{a^3}+\fr{c^4}{a^4}},
\ee
\ses
\be
s^3_{\{3\}}=
\fr
{
\fr{c^1}{a^1}-
\fr{c^2}{a^2}-\fr{c^3}{a^3}+\fr{c^4}{a^4}}
{
\fr{c^1}{a^1}+
\fr{c^2}{a^2}+\fr{c^3}{a^3}+\fr{c^4}{a^4}},
\ee
and
\be
s^1_{\{3\}}=
\fr
{
\fr{c^1}{b^1}\fr{b^1}{a^1}-
\fr{c^2}{b^2}\fr{b^2}{a^2}+\fr{c^3}{b^3}\fr{b^3}{a^3}-\fr{c^4}{b^4}\fr{b^4}{a^4}
}
{
\fr{c^1}{b^1}\fr{b^1}{a^1}+
\fr{c^2}{b^2}\fr{b^2}{a^2}+\fr{c^3}{b^3}\fr{b^3}{a^3}+\fr{c^4}{b^4}\fr{b^4}{a^4}
},
\ee
\ses
\be
s^2_{\{3\}}=
\fr
{
\fr{c^1}{b^1}\fr{b^1}{a^1}+
\fr{c^2}{b^2}\fr{b^2}{a^2}-\fr{c^3}{b^3}\fr{b^3}{a^3}-\fr{c^4}{b^4}\fr{b^4}{a^4}
}
{
\fr{c^1}{b^1}\fr{b^1}{a^1}+
\fr{c^2}{b^2}\fr{b^2}{a^2}+\fr{c^3}{b^3}\fr{b^3}{a^3}+\fr{c^4}{b^4}\fr{b^4}{a^4}
},
\ee
\ses
\be
s^3_{\{3\}}=
\fr
{
\fr{c^1}{b^1}\fr{b^1}{a^1}-
\fr{c^2}{b^2}\fr{b^2}{a^2}-\fr{c^3}{b^3}\fr{b^3}{a^3}+\fr{c^4}{b^4}\fr{b^4}{a^4}
}
{
\fr{c^1}{b^1}\fr{b^1}{a^1}+
\fr{c^2}{b^2}\fr{b^2}{a^2}+\fr{c^3}{b^3}\fr{b^3}{a^3}+\fr{c^4}{b^4}\fr{b^4}{a^4}
},
\ee
together with
\be
s^1_{\{3\}}=
\fr
{J_{11}J_{21}-J_{12}J_{22}+J_{13}J_{23}-J_{14}J_{24}}
{J_{11}J_{21}+J_{12}J_{22}+J_{13}J_{23}+J_{14}J_{24}},
\ee
\ses
\be
s^2_{\{3\}}=
\fr
{J_{11}J_{21}+J_{12}J_{22}-J_{13}J_{23}-J_{14}J_{24}}
{J_{11}J_{21}+J_{12}J_{22}+J_{13}J_{23}+J_{14}J_{24}},
\ee
\ses
\be
s^3_{\{3\}}=
\fr
{J_{11}J_{21}-J_{12}J_{22}-J_{13}J_{23}+J_{14}J_{24}}
{J_{11}J_{21}+J_{12}J_{22}+J_{13}J_{23}+J_{14}J_{24}}.
\ee
Due simplications yield eventually


\ses

PROPOSITION. The {\it $\cA^{\{+\}}_4$--composition law} (3.54) involves the explicit components
\be
s^1_{\{3\}}=
\fr
{s_{\{1\}}^1+s_{\{2\}}^1+s_{\{1\}}^2s_{\{2\}}^3+s_{\{1\}}^3s_{\{2\}}^2}
{1+s_{\{1\}}^1s_{\{2\}}^1
+s_{\{1\}}^2s_{\{2\}}^2
+s_{\{1\}}^3s_{\{2\}}^3},
\ee
\ses
\be
s^2_{\{3\}}=
\fr
{s_{\{1\}}^2
+s_{\{2\}}^2+s_{\{1\}}^1s_{\{2\}}^3+s_{\{1\}}^3s_{\{2\}}^1}
{1+s_{\{1\}}^1s_{\{2\}}^1
+s_{\{1\}}^2s_{\{2\}}^2+s_{\{1\}}^3s_{\{2\}}^3},
\ee
\ses
\be
s^3_{\{3\}}=
\fr
{s_{\{1\}}^3+s_{\{2\}}^3+s_{\{1\}}^1s_{\{2\}}^2+s_{\{1\}}^2s_{\{2\}}^1}
{1+s_{\{1\}}^1s_{\{2\}}^1+s_{\{1\}}^2s_{\{2\}}^2+s_{\{1\}}^3s_{\{2\}}^3}.
\ee

\ses

An alternative convenient way to obtain the formulae (3.64)--(3.66) is to resolve the subtraction law  set
(3.50)--(3.53)
with respect to the  quantities
$s^1_{\{3\}},\,s^2_{\{3\}}\,s^3_{\{3\}}$.


Realizing  the opposed way
\be
R^p_{\{b\}}=N^p_q(b,a)R^q_{\{a\}}
\ee
 from the reference frame $S\{a\}$
 to the reference frame $S\{b\}$ implies using the inverted coefficients
\be
N^p_q(b,a)=h^p_A(b)h^A_q(a).
\ee
We obtain
\be
N^0_0(b,a)=\fr{F(b)}{4F(a)}
\Bigl(
\fr{a^1}{b^1}+
\fr{a^2}{b^2}+\fr{a^3}{b^3}+\fr{a^4}{b^4}
\Bigr),
\ee
\ses
\be
N^1_0(b,a)=\fr{F(b)}{4F(a)}
\Bigl(
\fr{a^1}{b^1}-
\fr{a^2}{b^2}+\fr{a^3}{b^3}-\fr{a^4}{b^4}
\Bigr),
\ee
\ses\be
N^2_0(b,a)=\fr{F(b)}{4F(a)}
\Bigl(
\fr{a^1}{b^1}+
\fr{a^2}{b^2}-\fr{a^3}{b^3}-\fr{a^4}{b^4}
\Bigr),
\ee
\ses\be
N^3_0(b,a)=\fr{F(b)}{4F(a)}
\Bigl(
\fr{a^1}{b^1}-
\fr{a^2}{b^2}-\fr{a^3}{b^3}+\fr{a^4}{b^4}
\Bigr),
\ee
so that  the coefficients
$
N^p_q(b,a)
$
are obtainable from the above coefficients
$
N^p_q(a,b)
$
merely by substituting $a^A$ with $b^A$, and $b^A$ with $a^A$.


The  {\it reciprocal relative velocity }
\be
s^a(b,a)=\fr{N^a_0(b,a)}{N^0_0(b,a)}\equiv \fr{h^a_A(b)a^A}{h_B^0(b)a^B}
\ee
(cf. (3.17)) bears naturally the meaning of the velocity which is inverse to
the
$s^a(a,b)$:
\be
s^a(b,a)=\ominus
s^a(a,b).
\ee
Given the velocity $\{s^a\}$ with some   values
\be
s^1=s^1(a,b), \qquad
s^2=s^2(a,b), \qquad
s^3=s^3(a,b),
\ee
the calculation of
$\{s^a(b,a)\}$
leads us to the components
\be
\ominus s^1=
\fr14\Biggl[
\fr{1}{1+s^1+s^2+s^3}
-\fr{1}{1-s^1+s^2-s^3}
+\fr{1}{1+s^1-s^2-s^3}
-\fr{1}{1-s^1-s^2+s^3}
\Biggr],
\ee
\ses
\be
\ominus s^2=
\fr
14
\Biggl[
\fr{1}{1+s^1+s^2+s^3}
+\fr{1}{1-s^1+s^2-s^3}
-\fr{1}{1+s^1-s^2-s^3}
-\fr{1}{1-s^1-s^2+s^3}
\Biggr],
\ee
\ses
\be
\ominus s^3=
\fr
14
\Biggl[
\fr{1}{1+s^1+s^2+s^3}
-\fr{1}{1-s^1+s^2-s^3}
-\fr{1}{1+s^1-s^2-s^3}
+\fr{1}{1-s^1-s^2+s^3}
\Biggr],
\ee
or
\be
\ominus s^1=
-\fr1{(A(s))^4}
\Biggl[s^1-2s^2s^3-(s^1)^3+s^1(s^2s^2+s^3s^3)\Biggr],
\ee
\ses
\be
\ominus s^2=
-\fr1{(A(s))^4}
\Biggl[s^2-2s^1s^3-(s^2)^3+s^2(s^1s^1+s^3s^3)\Biggr],
\ee
\ses
\be
\ominus s^3=
-\fr1{(A(s))^4}
\Biggl[s^3-2s^1s^2-(s^3)^3+s^3(s^1s^1+s^2s^2)\Biggr].
\ee
\ses

PROPOSITION.
The  $\cA^{\{+\}}_4$--{\it reciprocity for three--dimensional velocities} reads as (3.79)--(3.81).

\ses


\ses

NOTE.
 From (3.50)--(3.66) the ordinary two--dimensional special--relativistic formulae ensue
as follows:
\be
\{s_{\{1\}}^2=s_{\{1\}}^3=s_{\{2\}}^2=s_{\{2\}}^3=0\}\to s_{\{1\}}^1=
\fr{s_{\{3\}}^1-s_{\{2\}}^1}{1-s_{\{2\}}^1s_{\{3\}}^1}
\quad
{\text{and}} \quad
s_{\{3\}}^1=\fr{s_{\{1\}}^1+s_{\{2\}}^1}{1+s_{\{1\}}^1s_{\{2\}}^1}.
\ee
In this case, the reciprocity of the ordinary ``obvious''  type
\be
\ominus s^a=-s^a
\ee
is a true implication from the law (3.79)--(3.81).

\ses


 In the small--velocity approximation up to O(5), we obtain
\be
A(s)\approx
A_1(s)+A_2(s)
\ee
with \be A_1(s)=1
-\fr12\Bigl(
(s^1)^2+(s^2)^2+(s^3)^2\Bigr)
-\fr18\Bigl(
(s^1)^4+(s^2)^4+(s^3)^4\Bigr)
\ee
and
\be
A_2(s)=2s^1s^2s^3
-\fr54\Bigl(
(s^1)^2(s^2)^2+(s^2)^2(s^3)^2+(s^1)^2(s^3)^2\Bigr).
\ee


\ses\ses

\setcounter{sctn}{4}
 \setcounter{equation}{0}

\nin{\bf 4. Invariance in $\cA_4^{\{+\}}$--space}

 \ses

Let us consider a  non--singular, and non--linear in general, transformation
\be
y^A=\cF^A(\wt y^B)
\ee
under which the Finslerian metric function remains invariant, that is,
\be
 F(y) =  F (\wt y).
 \ee
Let us construct from the coefficients
$
\cF^A
$
the derivatives
\be
\cF^A_B\eqdef\D{\cF^A}{\wt y^B}
\ee
and
\be
\cF^A_{BC}\eqdef\D{\cF^A_B}{\wt y^C}.
\ee
For our purposes it is worth assuming that the functions
${\cal F}^A$
are sufficiently smooth and positively homogeneous of degree 1 with respect to
$\wt y$,
so that
\be {\cal F}^A (k \wt y) = k {\cal F}^A(\wt
y), \qquad k > 0,
 \ee
 (for any admissible set of arguments).
The last condition guarantees retaining the homogeneity property for  the Finslerian metric function $F$
under the transformations (4.1)
and allows rewriting them in the form
\be
 y^A = {\cal F}^A_B(\wt y) \wt y^B
\ee
(as this immediately follows from the Euler theorem for homogeneous functions).
Generally speaking, the second derivatives
does not vanish identically:
\be
 {\cal F}^A_{BC} \ne 0.
 \ee

Differentiating (4.2) with respect to
 $\wt y^{C}$
 leads to  new identity
\be
\wt y_C = y_B {\cal F}^B_C,
 \ee
which in turn can be differentiated with respect to
$\wt y^D$,
which yields
\be
g_{CD} (\wt y) =
{\cal F}^A_{C}(\wt y) {\cal F}^B_D (\wt y) g_{AB} \left( {\cal
F}(\wt y) \right) + y_B {\cal F}^B_{CD}
 \ee
(the definition (1.3) has been used).

If the transformation
(4.1) fulfills also the condition
\be
 y_B{\cal F}^B_{C  D}= 0,
\ee
then we call it
{\it metric},
keeping in mind that  in such a case the transformation (4.9) leaves also
invariant the Finslerian metric tensor:
\be
g_{CD} (\wt y) =
{\cal F}^A_{C}(\wt y) {\cal F}^B_D (\wt y) g_{AB} \left( {\cal
F}(\wt y) \right).
 \ee
Owing to (1.6), the metricity condition (4.10) can be written as
\be
\cF_B\cF^B_{CD}\equiv 0
\ee
with the functions
 \be
 \cF_B=1/\cF^B.
\ee
Obviously, the metric transformations comprise s group.

\ses\ses

DEFINITION. Under the above conditions, the set of transformations (4.1)
 is called the {\it  group of Finslerian metric transformations}.

\ses

In case of the particular Finslerian metric function (1.5), an attentive consideration of the role of the
indicatrix variables
$\{u^a\}$ (see (1.39)) leads to the following conclusions.

\ses

PROPOSITION.
The Euclidean rotations of
the indicatrix variables $\{u^a\}$
give rise to the nonlinear transformations of the vectors
$\{y^A\}$,
which leave the Finslerian metric function (1.5) invariant and simultaneously realize invariance transformation (4.11)
for the associated Finslerian metric tensor.

\ses

The explicit form for the required coefficients  $\cF^A$ will be evaluated below in Appendix A.
Namely, under the rotation conditions (A.24)--(A.28), the nonlinear transformations under our present concern
prove to be given explicitly by means of the formulae (A.3)--(A.23). They involve three angles of rotations.
For the transformations obtained the validity of the metricity condition (4.10) can
be verified straightforwardly by applying the required Maple9-tools (see Appendix B below). The formulae are
essentially got simplified in case of one--angle--rotations (see Appendix C below).

Additionally, the translations in the indicatrix:
\be
\wt u^a=u^a+n^a
\ee
 induce obviously the unimodular dilatations
\be
\wt y^A=y^A\cdot k^A, \qquad k^1k^2k^3k^4=1,
\ee
because of the exponential nature of the indicatrix representation (1.39) of unit vectors.


\ses\ses

\setcounter{equation}{0}

{\bf Appendix A. Coefficients for three--angle rotations}

\ses

Let us start with an arbitrary linear nonsingular transformation of the indicatrix variables
 $\{u^a\}$ entering (1.39).
 Specifying them for definiteness to fulfill
 $ \ln l^1=\al+\beta+\ga,~ \ln l^2=-\al+\beta-\ga,~ \ln l^3=\al-\beta-\ga,~\ln l^4=-\al-\beta+\ga$
with $\{\al,\,\beta,\,\ga\}=\{u^1,u^2,u^3\}$,
 we have
\be
\al=
l_1\wt\al+l_2\wt\beta+l_3\wt\ga,
\qquad
\beta=
m_1\wt\al+m_2\wt\beta+m_3\wt\ga,
\qquad
\ga=
n_1\wt\al+n_2\wt\beta+n_3\wt\ga,
\ee
where
\be
\{m_1,m_2,m_3,n_1,n_2,n_3,l_1,l_2,l_3\}
\ee
is a set of constants.
This entails
$$
\ln a^1=
(l_1\wt\al+l_2\wt\beta+l_3\wt\ga)
+
(m_1\wt\al+m_2\wt\beta+m_3\wt\ga)
+
(n_1\wt\al+n_2\wt\beta+n_3\wt\ga),
$$
\ses
$$
\ln a^2=
-
(l_1\wt\al+l_2\wt\beta+l_3\wt\ga)
+
(m_1\wt\al+m_2\wt\beta+m_3\wt\ga)
-
(n_1\wt\al+n_2\wt\beta+n_3\wt\ga),
$$
\ses
$$
\ln a^3=
(l_1\wt\al+l_2\wt\beta+l_3\wt\ga)
-
(m_1\wt\al+m_2\wt\beta+m_3\wt\ga)
-
(n_1\wt\al+n_2\wt\beta+n_3\wt\ga),
$$
\ses
$$
\ln a^4=
-
(l_1\wt\al+l_2\wt\beta+l_3\wt\ga)
-
(m_1\wt\al+m_2\wt\beta+m_3\wt\ga)
+
(n_1\wt\al+n_2\wt\beta+n_3\wt\ga),
$$
or
$$
\ln a^1=
(l_1+m_1+n_1)\wt\al+
(l_2+m_2+n_2)\wt\beta+
(l_3+m_3+n_3)\wt\ga,
$$
\ses
$$
\ln a^2=
(-l_1+m_1-n_1)\wt\al+
(-l_2+m_2-n_2)\wt\beta+
(-l_3+m_3-n_3)\wt\ga,
$$
\ses
$$
\ln a^3=
(l_1-m_1-n_1)\wt\al+
(l_2-m_2-n_2)\wt\beta+
(l_3-m_3-n_3)\wt\ga,
$$
\ses
$$
\ln a^4=
(-l_1-m_1+n_1)\wt\al+
(-l_2-m_2+n_2)\wt\beta+
(-l_3-m_3+n_3)\wt\ga,
$$
from which it follows directly that
$$
4\ln a^1=
(l_1+m_1+n_1)
(\ln \wt a^1-\ln \wt a^2+\ln \wt a^3-\ln \wt a^4)
$$
\ses
$$
+
(l_2+m_2+n_2)
(\ln \wt a^1+\ln \wt a^2-\ln \wt a^3-\ln \wt a^4)
$$
\ses
$$
+
(l_3+m_3+n_3)
(\ln \wt a^1-\ln \wt a^2-\ln \wt a^3+\ln \wt a^4),
$$
\ses
\ses
$$
4\ln a^2=
(-l_1+m_1-n_1)
(\ln \wt a^1-\ln \wt a^2+\ln \wt a^3-\ln \wt a^4)
$$
\ses
$$
+
(-l_2+m_2-n_2)
(\ln \wt a^1+\ln \wt a^2-\ln \wt a^3-\ln \wt a^4)
$$
\ses
$$
+
(-l_3+m_3-n_3)
(\ln \wt a^1-\ln \wt a^2-\ln \wt a^3+\ln \wt a^4),
$$
\ses
\ses
$$
4\ln a^3=
(l_1-m_1-n_1)
(\ln \wt a^1-\ln \wt a^2+\ln \wt a^3-\ln \wt a^4)
$$
\ses
$$
+
(l_2-m_2-n_2)
(\ln \wt a^1+\ln \wt a^2-\ln \wt a^3-\ln \wt a^4)
$$
\ses
$$
+
(l_3-m_3-n_3)
(\ln \wt a^1-\ln \wt a^2-\ln \wt a^3+\ln \wt a^4),
$$
\ses
\ses
$$
4\ln a^4=
(-l_1-m_1+n_1)
(\ln \wt a^1-\ln \wt a^2+\ln \wt a^3-\ln \wt a^4)
$$
\ses
$$
+
(-l_2-m_2+n_2)
(\ln \wt a^1+\ln \wt a^2-\ln \wt a^3-\ln \wt a^4)
$$
\ses
$$
+
(-l_3-m_3+n_3)
(\ln \wt a^1-\ln \wt a^2-\ln \wt a^3+\ln \wt a^4),
$$
\ses
or
$$
4\ln a^1=
(l_1+m_1+n_1
+
l_2+m_2+n_2
+
l_3+m_3+n_3)
\ln \wt a^1
$$
\ses
$$
+
(-l_1-m_1-n_1
+
l_2+m_2+n_2
-
l_3-m_3-n_3)
\ln \wt a^2
$$
\ses
$$
+
(l_1+m_1+n_1
-
l_2-m_2-n_2
-
l_3-m_3-n_3)
\ln \wt a^3
$$
\ses
$$
+
(-l_1-m_1-n_1
-
l_2-m_2-n_2
+
l_3+m_3+n_3)
\ln \wt a^4,
$$
\ses
\ses
$$
4\ln a^2=
(-l_1+m_1-n_1
-
l_2+m_2-n_2
-
l_3+m_3-n_3)
\ln \wt a^1
$$
\ses
$$
+
(l_1-m_1+n_1
-
l_2+m_2-n_2
+
l_3-m_3+n_3)
\ln \wt a^2
$$
\ses
$$
+
(-l_1+m_1-n_1
+
l_2-m_2+n_2
+
l_3-m_3+n_3)
\ln \wt a^3
$$
\ses
$$
+
(l_1-m_1+n_1
+
l_2-m_2+n_2
-
l_3+m_3-n_3)
\ln \wt a^4,
$$
\ses
\ses
$$
4\ln a^3=
(l_1-m_1-n_1
+
l_2-m_2-n_2
+
l_3-m_3-n_3)
\ln \wt a^1
$$
\ses
$$
+
(-l_1+m_1+n_1
+
l_2-m_2-n_2
-
l_3+m_3+n_3)
\ln \wt a^2
$$
\ses
$$
+
(l_1-m_1-n_1
-
l_2+m_2+n_2
-
l_3+m_3+n_3)
\ln \wt a^3
$$
\ses
$$
+
(-l_1+m_1+n_1
-
l_2+m_2+n_2
+
l_3-m_3-n_3)
\ln \wt a^4,
$$
\ses
\ses
$$
4\ln a^4=
(-l_1-m_1+n_1
-
l_2-m_2+n_2
-
l_3-m_3+n_3)
\ln \wt a^1
$$
\ses
$$
+
(l_1+m_1-n_1
-
l_2-m_2+n_2
+
l_3+m_3-n_3)
\ln \wt a^2
$$
\ses
$$
+
(-l_1-m_1+n_1
+
l_2+m_2-n_2
+
l_3+m_3-n_3)
\ln \wt a^3
$$
\ses
$$
+
(l_1+m_1-n_1
+
l_2+m_2-n_2
-
l_3-m_3+n_3)
\ln \wt a^4.
$$

Thus we can conclude that
$$
(a^1)^4=
(\wt a^1)^
{
(l_1+m_1+n_1
+
l_2+m_2+n_2
+
l_3+m_3+n_3)
}
$$
\ses
$$
\cdot
( \wt a^2)^
{
(-l_1-m_1-n_1
+
l_2+m_2+n_2
-
l_3-m_3-n_3)
}
$$
\ses
$$
\cdot
(\wt a^3)^
{
(l_1+m_1+n_1
-
l_2-m_2-n_2
-
l_3-m_3-n_3)
}
$$
\ses
$$
\cdot
(\wt a^4)^
{
(-l_1-m_1-n_1
-
l_2-m_2-n_2
+
l_3+m_3+n_3)
},
$$
\ses
$$
(a^2)^4=
(\wt a^1)^
{
(-l_1+m_1-n_1
-
l_2+m_2-n_2
-
l_3+m_3-n_3)
}
$$
\ses
$$
\cdot
( \wt a^2)^
{
(l_1-m_1+n_1
-
l_2+m_2-n_2
+
l_3-m_3+n_3)
},
$$
\ses
$$
\cdot
(\wt a^3)^
{
(-l_1+m_1-n_1
+
l_2-m_2+n_2
+
l_3-m_3+n_3)
}
$$
\ses
$$
\cdot
(\wt a^4)^
{
(l_1-m_1+n_1
+
l_2-m_2+n_2
-
l_3+m_3-n_3)
},
$$
\ses
$$
(a^3)^4=
(\wt a^1)^
{
(l_1-m_1-n_1
+
l_2-m_2-n_2
+
l_3-m_3-n_3)
}
$$
\ses
$$
\cdot
( \wt a^2)^
{
(-l_1+m_1+n_1
+
l_2-m_2-n_2
-
l_3+m_3+n_3)
}
$$
\ses
$$
\cdot
(\wt a^3)^
{
(l_1-m_1-n_1
-
l_2+m_2+n_2
-
l_3+m_3+n_3)
}
$$
\ses
$$
\cdot
(\wt a^4)^
{
(-l_1+m_1+n_1
-
l_2+m_2+n_2
+
l_3-m_3-n_3)
},
$$
\ses
$$
(a^4)^4=
(\wt a^1)^
{
(-l_1-m_1+n_1
-
l_2-m_2+n_2
-
l_3-m_3+n_3)
}
$$
\ses
$$
\cdot
( \wt a^2)^
{
(l_1+m_1-n_1
-
l_2-m_2+n_2
+
l_3+m_3-n_3)
}
$$
\ses
$$
\cdot
(\wt a^3)^
{
(-l_1-m_1+n_1
+
l_2+m_2-n_2
+
l_3+m_3-n_3)
}
$$
\ses
$$
\cdot
(\wt a^4)^
{
(l_1+m_1-n_1
+
l_2+m_2-n_2
-
l_3-m_3+n_3).
}
$$

Also, we obtain
$$
(a^1a^2a^3)^4=
(\wt a^1)^
{
(l_1+m_1-n_1
+
l_2+m_2-n_2
+
l_3+m_3-n_3)
}
$$
\ses
$$
\cdot
( \wt a^2)^
{
(-l_1-m_1+n_1
+
l_2+m_2-n_2
-
l_3-m_3+n_3)
}
$$
\ses
$$
\cdot
(\wt a^3)^
{
(l_1+m_1-n_1
-
l_2-m_2+n_2
-
l_3-m_3+n_3)
}
$$
\ses
$$
\cdot
(\wt a^4)^
{
(-l_1-m_1+n_1
-
l_2-m_2+n_2
+
l_3+m_3-n_3)
},
$$
\ses
\ses
$$
(a^1a^2a^4)^4=
(\wt a^1)^
{
(-l_1+m_1+n_1
-
l_2+m_2+n_2
-
l_3+m_3+n_3)
}
$$
\ses
$$
\cdot
( \wt a^2)^
{
(l_1-m_1-n_1
-
l_2+m_2+n_2
+
l_3-m_3-n_3)
}
$$
\ses
$$
\cdot
(\wt a^3)^
{
(-l_1+m_1+n_1
+
l_2-m_2-n_2
+
l_3-m_3-n_3)
}
$$
\ses
$$
\cdot
(\wt a^4)^
{
(l_1-m_1-n_1
+
l_2-m_2-n_2
-
l_3+m_3+n_3)
},
$$
\ses
\ses
$$
(a^1a^3a^4)^4=
(\wt a^1)^
{
(l_1-m_1+n_1
+
l_2-m_2+n_2
+
l_3-m_3+n_3)
}
$$
\ses
$$
\cdot
( \wt a^2)^
{
(-l_1+m_1-n_1
+
l_2-m_2+n_2
-
l_3+m_3-n_3)
}
$$
\ses
$$
\cdot
(\wt a^3)^
{
(l_1-m_1+n_1
-
l_2+m_2-n_2
-
l_3+m_3-n_3)
}
$$
\ses
$$
\cdot
(\wt a^4)^
{
(l_1+m_1-n_1
-
l_2+m_2-n_2
+
l_3-m_3+n_3)
},
$$
\ses
\ses
$$
(a^2a^3a^4)^4=
(\wt a^1)^
{
(-l_1-m_1-n_1
-
l_2-m_2-n_2
-
l_3-m_3-n_3)
}
$$
\ses
$$
\cdot
( \wt a^2)^
{
(l_1+m_1+n_1
-
l_2-m_2-n_2
+
l_3+m_3+n_3)
}
$$
\ses
$$
\cdot
(\wt a^3)^
{
(-l_1-m_1-n_1
+
l_2+m_2+n_2
+
l_3+m_3+n_3)
}
$$
\ses
$$
\cdot
(\wt a^4)^
{
(l_1+m_1+n_1
+
l_2+m_2+n_2
-
l_3-m_3-n_3)
}.
$$
\ses

This way we get the coefficients of the transformation
\be
a^A=\cF^A(\wt a^B)
\ee
to explicitly read
\be
\cF^1=(\wt a^1)^{f^{11}}(\wt a^2)^{f^{12}}(\wt a^3)^{f^{13}}(\wt a^4)^{f^{14}},
\ee
\ses
\be
\cF^2=(\wt a^1)^{f^{21}}(\wt a^2)^{f^{22}}(\wt a^3)^{f^{23}}(\wt a^4)^{f^{24}},
\ee
\ses
\be
\cF^3=(\wt a^1)^{f^{31}}(\wt a^2)^{f^{32}}(\wt a^3)^{f^{33}}(\wt a^4)^{f^{34}},
\ee
\ses\be
\cF^4=(\wt a^1)^{f^{41}}(\wt a^2)^{f^{42}}(\wt a^3)^{f^{43}}(\wt a^4)^{f^{44}},
\ee
with
\ses
\be
f^{11}=
{
(l_1+m_1+n_1
+
l_2+m_2+n_2
+
l_3+m_3+n_3+1)/4
},
\ee
\ses
\be
f^{12}=
{
(-l_1-m_1-n_1
+
l_2+m_2+n_2
-
l_3-m_3-n_3+1)/4
},
\ee
\ses
\be
f^{13}=
{
(l_1+m_1+n_1
-
l_2-m_2-n_2
-
l_3-m_3-n_3+1)/4
},
\ee
\ses
\be
f^{14}=
{
(-l_1-m_1-n_1
-
l_2-m_2-n_2
+
l_3+m_3+n_3+1)/4
},
\ee
\ses
\ses
\be
f^{21}=
{
(-l_1+m_1-n_1
-
l_2+m_2-n_2
-
l_3+m_3-n_3+1)/4
},
\ee
\ses
\be
f^{22}=
{
(l_1-m_1+n_1
-
l_2+m_2-n_2
+
l_3-m_3+n_3+1)/4
},
\ee
\ses
\be
f^{22}=
{
(-l_1+m_1-n_1
+
l_2-m_2+n_2
+
l_3-m_3+n_3+1)/4
},
\ee
\ses
\be
f^{24}=
{
(l_1-m_1+n_1
+
l_2-m_2+n_2
-
l_3+m_3-n_3+1)/4
},
\ee
\ses
\ses
\be
f^{31}=
{
(l_1-m_1-n_1
+
l_2-m_2-n_2
+
l_3-m_3-n_3+1)/4
},
\ee
\ses
\be
f^{32}=
{
(-l_1+m_1+n_1
+
l_2-m_2-n_2
-
l_3+m_3+n_3+1)/4
},
\ee
\ses
\be
f^{33}=
{
(l_1-m_1-n_1
-
l_2+m_2+n_2
-
l_3+m_3+n_3+1)/4
}
\ee
\ses
\be
f^{34}=
{
(-l_1+m_1+n_1
-
l_2+m_2+n_2
+
l_3-m_3-n_3+1)/4
},
\ee
\ses
\ses
\be
f^{41}=
{
(-l_1-m_1+n_1
-
l_2-m_2+n_2
-
l_3-m_3+n_3+1)/4
},
\ee
\ses
\be
f^{42}=
{
(l_1+m_1-n_1
-
l_2-m_2+n_2
+
l_3+m_3-n_3+1)/4
},
\ee
\ses
\be
f^{43}=
{
(-l_1-m_1+n_1
+
l_2+m_2-n_2
+
l_3+m_3-n_3+1)/4
},
\ee
\ses
\be
f^{44}=
{
(l_1+m_1-n_1
+
l_2+m_2-n_2
-
l_3-m_3+n_3+1)/4
}.
\ee
\ses

Finally, we are to subject the coefficients (A.2) to the condition that
the transformation (A.1) realizes an euclidean rotation of the  set $\{\al,\beta,\ga\}$.
To this end it is convenient to accept
the  (Euler) three angle choice:
\be
c_1=\cos\theta,
\quad
c_2=\cos\psi,
\quad
c_3=\cos\phi,
\ee
\ses
\be
s_1=\sin\theta,
\quad
s_2=\sin\psi,
\quad
s_3=\sin\phi
\ee
to have
\ses
\be
l_1=c_2c_3-c_1s_2s_3,
\quad
m_1=s_2c_3+c_1c_2s_3,
\quad
n_1=s_1s_3,
\ee
\ses
\be
l_2=-c_2s_3-c_1s_2c_3,
\quad
m_2=-s_2s_3+c_1c_2c_3,
\quad
n_2=s_1c_3,
\ee
\ses
\be
l_3=s_1s_2,
\quad
m_3=-s_1c_2,
\quad
n_3=c_1.
\ee


\ses\ses

{\bf Appendix B. Three-angle.mws(by Maple9) }

\ses

The program presented below (created by means of Maple 9) does
evaluate the metricity condition which fulfillment means that the
transformation leaves the Finslerian metric tensor invariant.

\ses

\begin{verbatim}
> c1:=cos(theta);c2:=cos(psi);c3:=cos(phi);
> s1:=sin(theta);s2:=sin(psi);s3:=sin(phi);
> l1:=c2*c3-c1*s2*s3;l2:=-c2*s3-c1*s2*c3;l3:=s1*s2;
> m1:=s2*c3+c1*c2*s3;m2:=-s2*s3+c1*c2*c3;m3:=-s1*c2;
> n1:=s1*s3;n2:=s1*c3;n3:=c1;
\end{verbatim}

\[
 \mathit{c1} := \mathrm{cos}(\theta ); \quad
 \mathit{c2} := \mathrm{cos}(\psi ); \quad
 \mathit{c3} := \mathrm{cos}(\phi );
\]
\[
 \mathit{s1} := \mathrm{sin}(\theta ); \quad
 \mathit{s2} := \mathrm{sin}(\psi ); \quad
 \mathit{s3} := \mathrm{sin}(\phi );
\]
\[
 \mathit{l1} := \mathit{c2}\,\mathit{c3} - \mathit{c1}\,\mathit{s2}\,\mathit{s3}; \quad
 \mathit{l2} :=  - \mathit{c2}\,\mathit{s3} - \mathit{c1}\,\mathit{s2}\,\mathit{c3}; \quad
 \mathit{l3} := \mathit{s1}\,\mathit{s2}; \quad
\]
\[
 \mathit{m1} := \mathit{s2}\,\mathit{c3} + \mathit{c1}\,\mathit{c2}\,\mathit{s3}; \quad
 \mathit{m2} :=  - \mathit{s2}\,\mathit{s3} + \mathit{c1}\,\mathit{c2}\,\mathit{c3}; \quad
 \mathit{m3} :=  - \mathit{s1}\,\mathit{c2};
\]
\[
 \mathit{n1} := \mathit{s1}\,\mathit{s3}; \quad
 \mathit{n2} := \mathit{s1}\,\mathit{c3}; \quad
 \mathit{n3} := \mathit{c1}
\]

\begin{verbatim}
> F1:=(e1)^((l1+m1+n1+l2+m2+n2+l3+m3+n3+1)/4)*
      (e2)^((-l1-m1-n1+l2+m2+n2-l3-m3-n3+1)/4)*
      (e3)^((l1+m1+n1-l2-m2-n2-l3-m3-n3+1)/4)*
      (e4)^((-l1-m1-n1-l2-m2-n2+l3+m3+n3+1)/4):
\end{verbatim}

\begin{verbatim}
> F2:=(e1)^((-l1+m1-n1-l2+m2-n2-l3+m3-n3+1)/4)*
      (e2)^((l1-m1+n1-l2+m2-n2+l3-m3+n3+1)/4)*
      (e3)^((-l1+m1-n1+l2-m2+n2+l3-m3+n3+1)/4)*
      (e4)^((l1-m1+n1+l2-m2+n2-l3+m3-n3+1)/4):
\end{verbatim}

\begin{verbatim}
> F3:=(e1)^((l1-m1-n1+l2-m2-n2+l3-m3-n3+1)/4)*
      (e2)^((-l1+m1+n1+l2-m2-n2-l3+m3+n3+1)/4)*
      (e3)^((l1-m1-n1-l2+m2+n2-l3+m3+n3+1)/4)*
      (e4)^((-l1+m1+n1-l2+m2+n2+l3-m3-n3+1)/4):
\end{verbatim}

\begin{verbatim}
> F4:=(e1)^((-l1-m1+n1-l2-m2+n2-l3-m3+n3+1)/4)*
      (e2)^((l1+m1-n1-l2-m2+n2+l3+m3-n3+1)/4)*
      (e3)^((-l1-m1+n1+l2+m2-n2+l3+m3-n3+1)/4)*
      (e4)^((l1+m1-n1+l2+m2-n2-l3-m3+n3+1)/4):
\end{verbatim}

\begin{verbatim}
> a:=array(1..4,1..4):
> for i from 1 to 4
  do
     for j from 1 to 4
     do
        a[i,j]:=diff(F||i,e||j);
     end do:
  end do:
\end{verbatim}

\begin{verbatim}
> b:=array(1..4,1..4):
> for i from 1 to 4
  do
     for j from 1 to 4
     do
        b[i,j]:=simplify(add(1/F||k*diff(a[k,i],e||j),k=1..4),symbolic);
     end do:
  end do:
\end{verbatim}

\begin{verbatim}
> print(b);
\end{verbatim}

\[
 \left[
{\begin{array}{rrrr}
0 & 0 & 0 & 0 \\
0 & 0 & 0 & 0 \\
0 & 0 & 0 & 0 \\
0 & 0 & 0 & 0
\end{array}}
 \right]
\]

\ses

The result that all the entries of the matrix are zeroes means
that the metricity condition holds true.


\ses\ses

\setcounter{equation}{0}

{\bf Appendix C. One--angle rotation}

\ses\ses

Let us take the particular case
\be
\al=\wt\al\cos\eta+\wt\beta\sin\eta,
\qquad
\beta=-\wt\al\sin\eta+\wt\beta\cos\eta,
\qquad
\ga=\wt\ga
\ee
which represents the rotation by one angle, $\eta$, in the $\ga$--plane.
We get
\ses
$$
\ln a^1=\wt\al\cos\eta+\wt\beta\sin\eta
-\wt\al\sin\eta+\wt\beta\cos\eta+\wt\ga,
$$
\ses
\ses
$$
\ln a^2=-\wt\al\cos\eta-\wt\beta\sin\eta
-\wt\al\sin\eta+\wt\beta\cos\eta-\wt\ga,
$$
\ses
\ses
$$
\ln a^3=\wt\al\cos\eta+\wt\beta\sin\eta+\wt\al\sin\eta-\wt\beta\cos\eta-\wt\ga,
$$
\ses
\ses
$$
\ln a^4=-\wt\al\cos\eta-\wt\beta\sin\eta+\wt\al\sin\eta-\wt\beta\cos\eta+\wt\ga,
$$
\ses
or
$$
\ln a^1=\wt\al(\cos\eta-\sin\eta)
+\wt\beta(\cos\eta+\sin\eta)+\wt\ga,
$$
\ses
\ses
$$
\ln a^2=-\wt\al(\cos\eta+\sin\eta)
+\wt\beta(\cos\eta-\sin\eta)-\wt\ga,
$$
\ses
\ses
$$
\ln a^3=\wt\al(\cos\eta+\sin\eta)
-\wt\beta(\cos\eta-\sin\eta)-\wt\ga,
$$
\ses
\ses
$$
\ln a^4=-\wt\al(\cos\eta-\sin\eta)
-\wt\beta(\cos\eta+\sin\eta)+\wt\ga,
$$
which entails
\ses\\
$$
4\ln a^1=
(\ln \wt a^1-\ln \wt a^2+\ln \wt a^3-\ln \wt a^4)
(\cos\eta-\sin\eta)
$$
\ses
$$
+
(\ln \wt a^1+\ln \wt a^2-\ln \wt a^3-\ln \wt a^4)
(\cos\eta+\sin\eta)
$$
\ses
$$
+
(\ln \wt a^1-\ln \wt a^2-\ln \wt a^3+\ln \wt a^4),
$$
\ses
\ses
$$
4\ln a^2=-
(\ln \wt a^1-\ln \wt a^2+\ln \wt a^3-\ln \wt a^4)
(\cos\eta+\sin\eta)
$$
\ses
$$
+
(\ln \wt a^1+\ln \wt a^2-\ln \wt a^3-\ln \wt a^4)
(\cos\eta-\sin\eta)
$$
\ses
$$
-
(\ln \wt a^1-\ln \wt a^2-\ln \wt a^3+\ln \wt a^4),
$$
\ses
\ses
$$
4\ln a^3=
(\ln \wt a^1-\ln \wt a^2+\ln \wt a^3-\ln \wt a^4)
(\cos\eta+\sin\eta)
$$
\ses
$$
-
(\ln \wt a^1+\ln \wt a^2-\ln \wt a^3-\ln \wt a^4)
(\cos\eta-\sin\eta)
$$
\ses
$$
-
(\ln \wt a^1-\ln \wt a^2-\ln \wt a^3+\ln \wt a^4),
$$
\ses
\ses
$$
4\ln a^4=
-
(\ln \wt a^1-\ln \wt a^2+\ln \wt a^3-\ln \wt a^4)
(\cos\eta-\sin\eta)
$$
\ses
$$
-
(\ln \wt a^1+\ln \wt a^2-\ln \wt a^3-\ln \wt a^4)
(\cos\eta+\sin\eta)
$$
\ses
$$
+
(\ln \wt a^1-\ln \wt a^2-\ln \wt a^3+\ln \wt a^4),
$$
\ses
from which it follows that
$$
4\ln a^1=
(\ln \wt a^1-\ln \wt a^2+\ln \wt a^3-\ln \wt a^4)
\cos\eta
$$
\ses
$$
+
(\ln \wt a^1+\ln \wt a^2-\ln \wt a^3-\ln \wt a^4)
\cos\eta
$$
\ses
\ses
$$
-(\ln \wt a^1-\ln \wt a^2+\ln \wt a^3-\ln \wt a^4)
\sin\eta
$$
\ses
$$
+
(\ln \wt a^1+\ln \wt a^2-\ln \wt a^3-\ln \wt a^4)
\sin\eta
$$
\ses
$$
+
(\ln \wt a^1-\ln \wt a^2-\ln \wt a^3+\ln \wt a^4),
$$
\ses
\ses
so that
$$
4\ln a^1=
2(\ln \wt a^1-\ln \wt a^4)
\cos\eta
$$
\ses
\ses
$$
+2(\ln \wt a^2-\ln \wt a^3)
\sin\eta
$$
\ses
$$
+
\ln \wt a^1-\ln \wt a^2-\ln \wt a^3+\ln \wt a^4,
$$
\ses
$$
4\ln a^2=-
(\ln \wt a^1-\ln \wt a^2+\ln \wt a^3-\ln \wt a^4)
\cos\eta
$$
\ses
$$
+
(\ln \wt a^1+\ln \wt a^2-\ln \wt a^3-\ln \wt a^4)
\cos\eta
$$
\ses
\ses
$$
-
(\ln \wt a^1-\ln \wt a^2+\ln \wt a^3-\ln \wt a^4)
\sin\eta
$$
\ses
$$
-
(\ln \wt a^1+\ln \wt a^2-\ln \wt a^3-\ln \wt a^4)
\sin\eta
$$
\ses
\ses
$$
-(\ln \wt a^1-\ln \wt a^2-\ln \wt a^3+\ln \wt a^4),
$$
\ses
so that
$$
4\ln a^2=
2(\ln \wt a^2-\ln \wt a^3)
\cos\eta
$$
\ses
\ses
$$
-
2(\ln \wt a^1-\ln \wt a^4)
\sin\eta
$$
\ses
\ses
$$
-(\ln \wt a^1-\ln \wt a^2-\ln \wt a^3+\ln \wt a^4),
$$
\ses
$$
4\ln a^3=
(\ln \wt a^1-\ln \wt a^2+\ln \wt a^3-\ln \wt a^4)
\cos\eta
$$
\ses
$$
-
(\ln \wt a^1+\ln \wt a^2-\ln \wt a^3-\ln \wt a^4)
\cos\eta
$$
\ses
\ses
$$
+
(\ln \wt a^1-\ln \wt a^2+\ln \wt a^3-\ln \wt a^4)
\sin\eta
$$
\ses
$$
+
(\ln \wt a^1+\ln \wt a^2-\ln \wt a^3-\ln \wt a^4)
\sin\eta
$$
\ses
$$
-
(\ln \wt a^1-\ln \wt a^2-\ln \wt a^3+\ln \wt a^4),
$$
\ses
so that
$$
4\ln a^3=
-2(\ln \wt a^2-\ln \wt a^3)
\cos\eta
$$
\ses
\ses
$$
+
2(\ln \wt a^1-\ln \wt a^4)
\sin\eta
$$
\ses
\ses
$$
-
(\ln \wt a^1-\ln \wt a^2-\ln \wt a^3+\ln \wt a^4),
$$
\ses
$$
4\ln a^4=
-
(\ln \wt a^1-\ln \wt a^2+\ln \wt a^3-\ln \wt a^4)
\cos\eta
$$
\ses
$$
-
(\ln \wt a^1+\ln \wt a^2-\ln \wt a^3-\ln \wt a^4)
\cos\eta
$$
\ses
\ses
$$
+
(\ln \wt a^1-\ln \wt a^2+\ln \wt a^3-\ln \wt a^4)
\sin\eta
$$
\ses
$$
-
(\ln \wt a^1+\ln \wt a^2-\ln \wt a^3-\ln \wt a^4)
\sin\eta
$$
\ses
$$
+
(\ln \wt a^1-\ln \wt a^2-\ln \wt a^3+\ln \wt a^4),
$$
\ses
so that
$$
4\ln a^4=
-
2(\ln \wt a^1-\ln \wt a^4)
\cos\eta
-
2(\ln \wt a^2-\ln \wt a^3)
\sin\eta
$$
\ses
\ses
$$
+
(\ln \wt a^1-\ln \wt a^2-\ln \wt a^3+\ln \wt a^4).
$$
Eventually we obtain
\be
4\ln a^1=
2(\ln \wt a^1-\ln \wt a^4)
\cos\eta
+2(\ln \wt a^2-\ln \wt a^3)
\sin\eta
+
\ln \wt a^1-\ln \wt a^2-\ln \wt a^3+\ln \wt a^4,
\ee
\ses
\ses
\be
4\ln a^2=
2(\ln \wt a^2-\ln \wt a^3)
\cos\eta
-
2(\ln \wt a^1-\ln \wt a^4)
\sin\eta
-(\ln \wt a^1-\ln \wt a^2-\ln \wt a^3+\ln \wt a^4),
\ee
\ses
\ses
\be
4\ln a^3=
-2(\ln \wt a^2-\ln \wt a^3)
\cos\eta
+
2(\ln \wt a^1-\ln \wt a^4)
\sin\eta
-
(\ln \wt a^1-\ln \wt a^2-\ln \wt a^3+\ln \wt a^4),
\ee
\ses
\ses
\be
4\ln a^4=
-
2(\ln \wt a^1-\ln \wt a^4)
\cos\eta
-
2(\ln \wt a^2-\ln \wt a^3)
\sin\eta
+
(\ln \wt a^1-\ln \wt a^2-\ln \wt a^3+\ln \wt a^4).
\ee

The respective generalized rotation coefficients are given by the list:
\be
\cF^1=
(\wt a^1)^
{
(2\cos\eta+1)/4
}
( \wt a^2)^
{
(2\sin\eta-1)/4
}
(\wt a^3)^
{
(-2\sin\eta-1)/4
}
(\wt a^4)^
{
(-2\cos\eta+1)/4
},
\ee
\ses
\be
\cF^2=
(\wt a^1)^
{
(-2\sin\eta-1)/4
}
( \wt a^2)^
{
(2\cos\eta+1)/4
}
(\wt a^3)^
{
(-2\cos\eta+1)/4
}
(\wt a^4)^
{
(2\sin\eta-1)/4
},
\ee
\ses
\be
\cF^3=
(\wt a^1)^
{
(2\sin\eta-1)/4
}
( \wt a^2)^
{
(-2\cos\eta+1)/4
}
(\wt a^3)^
{
(2\cos\eta+1)/4
}
(\wt a^4)^
{
(-2\sin\eta-1)/4
},
\ee
\ses
\be
\cF^4=
(\wt a^1)^
{
(-2\cos\eta+1)/4
}
( \wt a^2)^
{
(-2\sin\eta-1)/4
}
(\wt a^3)^
{
(2\sin\eta-1)/4
}
(\wt a^4)^
{
(2\cos\eta+1)/4
}.
\ee


\ses\ses

{\bf Discussion}

\ses

 The $(N=2)$--dimensional precursor of the space (1.2)
 under our study is the
ordinary hyperbolic relativistic space
$
\cA_2~:=\{V_2,e_1,e_2,F^{\{two-dimensional\}}(y)\}
$
with
 $
F^{\{two-dimensional\}}=\sqrt{|y^1y^2|}\equiv{|t^2-x^2|},
 $
where
$
 t=(y^1+y^2)/2, ~ x=(y^1-y^2)/2.
 $
The anisotropic method exposed in the previous sections to increase
 the dimension $N$ and arrive at the  $\cA_4^{\{+\}}$--space differs  drastically from the isotropic
conventional  pseudoeuclidean way. Generally, our methods  of analysis were founded
 upon usage of the indicatrix geometry and indicatrix coordinates.
The conformal nature of the associated Finslerian metric tensor (exhibited by (1.53))
has played  a crucial role in Section 2 in our getting explicit solutions
to the  $\cA_4^{\{+\}}$--geodesic equations. They are that solutions that entailed the
 distance, angle, and scalar product for the
$\cA_4^{\{+\}}$--space.

In Section 3  the substantive items were concerned  the  $\cA_4^{\{+\}}$--kinematic aspects.
 Actually, all the obtained kinematic implications, included the
 $\cA_4^{\{+\}}$--extension (1.59)--(1.62) of the Lorentz transformations and the
 $\cA_4^{\{+\}}$--extension (1.63) of the pseudoeuclidean kinematic length of vector,
  as well  the composition and subtraction laws found in Section 3,
  were direct implications of the particular
 $\cA_4^{\{+\}}$--tetrad
$h^p_A$ (introduced in Section 1 by means of   (1.51)).
Since the tetrads present geometrically the reference systems proper for the inertial reference frames
(according to general kinematic principles; see, e.g., [3-5]), and the vector components
$\{Y^p\}$ entering the representations (1.59)--(1.63) are the tetradic components, the representations
bear precisely physical space--time meaning; namely,
$\{Y^0\}$
and
$\{Y^1,Y^2,Y^3\}$ are respectively the time components and the spatial components of the vector
$\{Y^p\}$ as observed in the inertial reference frame. Similar Finslerian kinematic ideas were
applied in context of the Finsleroid theory [6-10].

Calculations involved the constants $C^A_p$ coming
 from the fundamental indicatrix representation
(1.39) of the unit vector $l^A$.
When dealing  with the four--dimensional case, the following suitable choice
 $
C^A_0=\{1,1,1,1\},~ C^A_1=\{-\sqrt3,\, 1/{\sqrt3},\,
1/{\sqrt3},\, 1/{\sqrt3}\},
~   C^A_2=\{0,\,
\sqrt{8/3},\, -\sqrt{2/3},\, -\sqrt{2/3}\}, ~
 C^A_3=\{0,\,0,\,-\sqrt2,\,\sqrt2\}
$
was proposed in [1];
the  possibility
$
 C^A_0=\{1,1,1,1\}, ~
C^A_1=\{1,-1,1,-1\},
~  C^A_2=\{1,1,-1,-1\}, ~
C^A_3=\{1,-1,-1,1\}
$
which does not involve  any roots was
used in [11]. Obviously, these two sets of constants may be
expressed one through another by means of due euclidean rotations. However,
 many fundamental implications, e.g. the angle (2.48), are independent of any such choice.
With the latter choice of the constants,
the tetradic components gain the particularly simple structure, namely,
$$
h^0_A=l_A=\fr F4\lf(\fr1{a^1}, \fr1{a^2},\fr1{a^3},\fr1{a^4}\rg),
\qquad
h^1_A=\fr F4\lf(\fr1{a^1}, - \fr1{a^2},\fr1{a^3},-\fr1{a^4}\rg),
$$
\ses
$$
h^2_A=\fr F4\lf(\fr1{a^1}, \fr1{a^2},-\fr1{a^3},-\fr1{a^4}\rg),
\qquad
h^3_A=\fr F4\lf(\fr1{a^1},- \fr1{a^2},-\fr1{a^3},\fr1{a^4}\rg),
$$
\ses
together with their reciprocal components
\ses
$$
h_0^A=l^A=\fr1F\lf(a^1, a^2,a^3,a^4\rg),
\qquad
h_1^A=\fr1F\lf(a^1, -a^2,a^3,-a^4\rg),
$$
\ses
$$
h_2^A=\fr1F\lf(a^1, a^2,-a^3,-a^4\rg),
\qquad
h_3^A=\fr1F\lf(a^1, -a^2,-a^3,a^4\rg).
$$
\ses
The
Finslerian metric tensor, which components can be presented by
$$
g_{AB}=h^0_Ah^0_B-h^1_Ah^1_B-h^2_Ah^2_B-h^3_Ah^3_B,\qquad
g^{AB}=h_0^Ah_0^B-h_1^Ah_1^B-h_2^Ah_2^B-h_3^Ah_3^B,
$$
discribes the Finslerian aspects of geometry of the
$\cA_4^{\{+\}}$--space.

Study of the invariance properties of the
$\cA_4^{\{+\}}$--space faced us to conclude in Section 4
that the associated group of invariance
is a nonlinear representation of the euclidean group of
rotations and translations given rise to by the
induced euclidean structure of the generalized
$\cA_4^{\{+\}}$--hyperboloid (which is the indicatrix of the space under study).
At the same time, the
$\cA_4^{\{+\}}$--kinematic invariance transformations (1.59)--(1.64)
 are of the essentially linear structure
entailed by the tetrad projections.

Recently, a new  impetus to development of geometrical and relativistic applications
on the basis of the Berwald--Moor metric function was given in the work [11].
An intrinsic relationship with the hypercomplex numbers was emphasized, therefore
 the space was denoted by $H_4$.
The work was motivated by the desire to develop the theory prolonged
``beyond square--root concepts''.
It was argued that the ordinary physical motivation
that only the Minkowskian framework
 with   its   well--accepted characteristics can be the
basis proper to the nowadays theoretical and cosmological
real world models is probably not exactly true.
A novel feature was indicated that,  while
in the Minkowskian space the set of points equidistant
from two static events (the space of  relatively simultaneous
events) forms a hyperplane,  in the $H_4$--space the set probably forms a
non-linear surface.
Interesting list of   Research Problems was set forth in [11]
to investigate possibilities of finding
adequate metric quantities, angles and scalar products included,
the required conformal and congruent transformations as well as
 appropriate extended $H_4$--rotation transformations.

\ses

Generally, search for novel generalized physical aspects produced tentatively  by the
anisotropic  structure of the space--time,  in particular those referred
to light behaviour, seems to be an urgent task for the new Finslerian framework
outlined above.
We hope, in particular,  that the
$\cA_4^{\{+\}}$-- extension (1.59)--(1.62) of the Lorentz transformations
may serve to propose the  necessary kinematic grounds to think of respective light experiments.


\ses
\ses

\def\bibit[#1]#2\par{\rm\noindent\parskip1pt
                     \parbox[t]{.05\textwidth}{\mbox{}\hfill[#1]}\hfill
                     \parbox[t]{.925\textwidth}{\baselineskip11pt#2}\par}

\nin
{\bf References}
\bigskip

\bibit[1] G. S. Asanov: \it Finsler Geometry, Relativity and Gauge Theories, \rm
D.~Reidel Publ. Comp., Dordrecht 1985.

\bibit[2] H.~ Rund: \it The Differential Geometry of Finsler spaces, \rm
Springer-Verlag, Berlin 1959.

\bibit[3]  J. L. Synge, \it Relativity: the General Theory
\rm (North-Holland, Amsterdam, 1960).

\bibit[4] H.-J. Treder, H.-H. von Borzeszkowski, A. van der Merwe, and W~Yourgrau,\\
\it Fundamental Principles of General Relativity Theories. Local and Global Aspects\\
 of  Gravitation and Cosmology \rm (Plenum, N.\,Y., 1980).

\bibit[5]  M. Friedman, \it Foundations of Space-Time Theories
\rm
(Princeton University Press, Princeton, 1983).

\bibit[6] G.S. Asanov, ``The Finsler-type recasting of Lorentz transformations."
In: Proceedings  of Conference  {\it Physical Interpretation of
Relativity Theory}, September 15-20 (London, Sunderland, 2000), pp. 1-24.

\bibit[7] G.S. Asanov, \it Found. Phys. Lett. \bf15\rm, 199 (2002).

\bibit[8] G.S.~ Asanov: \it Rep. Math. Phys. \bf 45 \rm(2000), 155;
\bf 47 \rm(2001), 323.

\bibit[9] G.S.~ Asanov:
arXiv:hep-ph/0306023, 2003;
arXiv:math-ph/0310019, 2003;
arXiv:math.MG/0402013, 2004.

\bibit[10] G. S. Asanov, Finsleroid-Space,   arXiv:math-ph/0406029.

 \bibit[11] D.G. Pavlov:
Hypercomplex Numbers, Associated Metric Spaces,
and Extension of
Relativistic Hyperboloid,
 gr-qc/0206004.

\end{document}